\documentclass[useAMS,usenatbib]{mn2e}
\usepackage{amsmath,amsfonts,bbm,graphicx}
\usepackage{aas_macros}
\usepackage{url}
\usepackage{multirow}
\def\be{\begin{equation}}
\def\ee{\end{equation}}
\def\ba{\begin{eqnarray}}
\def\ea{\end{eqnarray}}

\newcommand{\vmax}{$v_{\mathrm{max}}\,$}
\newcommand{\vmpeak}{$v_{m_{peak}}\,$}
\newcommand{\mpeak}{$m_{{peak}}\,$}
\newcommand{\vpeak}{$v_{{peak}}\,$}
\newcommand{\hGpc}{Gpc~h$^{-1}$}

\usepackage[usenames]{color}

\def\<{\langle}
\def\>{\rangle}

\title[Disentangling redshift-space distortions and nonlinear bias ]{Disentangling redshift-space distortions and nonlinear bias using the 2D power
spectrum} %
\author[E. Jennings et al]{ Elise Jennings$^{1,2}$\thanks{E-mail: elise@fnal.gov}, Risa H. Wechsler$^{3,4}$, Samuel W. Skillman$^{3}$, Michael S. Warren$^{5}$\\ 
$^{1}$Center for Particle Astrophysics, Fermi National Accelerator Laboratory MS209, P.O. Box 500, Kirk Rd. \& Pine St., Batavia, IL 60510-0500\\
$^{2}$Kavli Institute for Cosmological Physics, Enrico Fermi Institute, University of Chicago, Chicago, IL 60637\\
$^{3}$ Kavli Institute for Particle Astrophysics and Cosmology,
Department of Physics, Stanford University, Stanford, CA, 94305\\
$^{4}$ SLAC National Accelerator Laboratory, Menlo Park, CA, 94025\\
$^{5}$Theoretical Division, LANL, Los Alamos, NM 87545\\
}

\begin{document}

\date{}

%\pagerange{\pageref{firstpage}--\pageref{lastpage}} \pubyear{2015}

\maketitle

%\label{firstpage}

\begin{abstract}
We present the 2D redshift space galaxy power spectrum, $P(k,\mu)$, measured from the Dark Sky simulations, using catalogs constructed with halo occupation distribution and subhalo abundance matching methods, chosen to represent an intermediate redshift sample of luminous red galaxies. We find that the information content in individual $\mu$ (cosine of the angle to the line of sight) bins is substantially richer then multipole moments, and show that this can be used to isolate the impact of nonlinear growth and redshift space distortion (RSD) effects. Using the $\mu<0.2$ simulation data, which is not impacted by RSD, we can successfully measure the nonlinear bias to $\sim 5$\% at $k<0.6 h$Mpc$^{-1}$.  Using the low $\mu$ simulation data to constrain the nonlinear bias, and $\mu\ge0.2$ to constrain the growth rate, we show that $f$ can be constrained to $\sim 26 (22)$\% to a $k_{\rm max}< 0.4 (0.6) h$Mpc$^{-1}$ from clustering alone using a dispersion model, for a range of galaxy models.  Our analysis of individual $\mu $ bins reveals interesting physical effects which arise from different methods of populating halos with galaxies. We find a prominent turnaround scale, at which RSD damping effects are greater than the nonlinear growth, which differs for each galaxy model. The idea of separating nonlinear growth and RSD effects making use of the full information in the 2D galaxy power spectrum yields significant improvements in constraining cosmological parameters and may be a promising probe of galaxy formation models.
\end{abstract}

\section{Introduction}
Peculiar velocity flows distort the large-scale mass distribution on Mpc scales in the Universe and are a fundamental cosmological observable that allows us to constrain key parameters of the  $\Lambda$CDM model and to look for deviations from this standard model. One of the key aims of future galaxy redshift surveys, e.g. 
Euclid \citep{Cimatti:2009is}, WFIRST \citep{2013arXiv1305.5422S} and the Dark Energy Spectroscopic Instrument (DESI) survey \citep{2013arXiv1308.0847L, 2015AAS...22533605E}
is to measure the linear perturbation theory relation between the density and velocity fields, referred to as the linear growth rate, to roughly 1\% precision using the redshift-space clustering statistics of different galaxy tracers. It is usual to study the multipole moments, either the monopole or quadrupole, of the power spectrum in redshift space, where peculiar velocities distort the clustering signal along the line of sight, which involves integrating out the $\mu$ dependence. In this paper we examine the full 2D power spectrum, $P(k,\mu)$  in order to isolate the impact of nonlinear growth and redshift-space distortion (RSD) effects. We use state-of-the-art simulations to generate mock galaxy samples with a variety of assumptions for how galaxies populate halos and compare the redshift-space clustering signals in each.

Dark matter halos are collapsed virialized structures which create deep potential wells in which galaxies are expected to reside. As a result, galaxies are biased tracers of the underlying dark matter, and their relative clustering signals have a non-trivial scale-dependent relation, often referred to as the nonlinear bias. 
A common approach to understanding this galaxy--halo connection is to use a halo occupation distribution (HOD) \citep[e.g.][]{2002ApJ...575..587B, 2002MNRAS.329..246B, 2005ApJ...633..791Z}, which models the probability that a halo of fixed virial mass hosts a certain number of galaxies, and to then constrain the parameters of this relationship using measurements such as the projected galaxy two-point clustering signal. Another approach, which takes into account mergers and the dependence of clustering on mass accretion histories, is to use abundance matching between galaxies and dark matter halos in simulations \citep{2004ApJ...609...35K,2004MNRAS.353..189V, 2006ApJ...647..201C}.  Precise models of how galaxies populate halos and how this connection may evolve with time are important for constraining galaxy formation scenarios. 

Current models for the two-point clustering statistics in redshift space that include perturbation theory expansions have been shown to be an improvement over linear theory in modeling redshift-space clustering statistics. Although all are limited to very large scales $k<0.15h$Mpc$^{-1}$ at low redshifts \citep[see e.g][]{Scoccimarro:2004tg, 2011MNRAS.410.2081J, 2012ApJ...748...78K} and moreover may only apply to highly biased tracers \citep{2011MNRAS.417.1913R}; none of the models can recover the linear growth rate to percent level accuracy on the scales that will be probed by future galaxy surveys. One of the key degeneracies in accurately constraining the growth rate is the nonlinear scale-dependent bias between the galaxies and dark matter. 

To distinguish between competing explanations for the accelerating expansion of the Universe, we need to measure the growth of structure to an accuracy of a few percent over a wide redshift interval. The next generation of galaxy redshift surveys, such as the Dark Energy Spectroscopic Instrument (DESI) \citep{2015AAS...22533605E}, will be able to achieve this precision.  One of the main aims of DESI is to measure luminous red galaxies (LRGs) up to $z = 1.0$, extending the BOSS LRG survey \cite[e.g.][]{2014MNRAS.443.1065B} in both redshift and survey area. In this paper we present predictions for the redshift-space power spectrum of several mock LRG samples at a number density and redshift relevant for DESI.  We analyze different $\mu$ bins and demonstrate how these can be used to isolate the impact of nonlinear growth and RSD. We note that the methods described here should also be applicable to galaxy samples selected from other surveys and with other selection techniques.

It is important to understand the sensitivity of these results to uncertainties in the galaxy--halo connection.  However, to date there have been relatively few studies of the redshift-space clustering signal using different galaxy models. Using a HOD to populate dark matter halos, \citet{2006MNRAS.368...85T} analyzed the redshift-space clustering signal in a simulation box of 253 $h^{-1}$Mpc on a side, but did not recover the linear theory predictions on scales accessible to their simulations. More recently, \citet{2015arXiv150303973Y} presented redshift-space clustering results using subhalo abundance matching methods in computational boxes of $300h^{-1}$Mpc on a side with 1024$^3$ particles, but do not present linear theory predictions which would demonstrate convergence of the RSD signal on  large scales. A key criterion for a robust redshift-space distortion analysis is both high mass resolution, for accurate velocity statistics, as well as a large computation volume to recover linear theory predictions on large scales \citep[][]{2015MNRAS.446..793J}.  If subhalo abundance matching is used to populate the simulation with galaxies, this places even stricter constraints on both the mass and force resolution required to resolve substructure. Here we use a new $1h^{-1}$Gpc cosmological box with 10240$^3$ particles from the {\em Dark Sky} series \citep{2014arXiv1407.2600S} to create both HOD and subhalo abundance matching catalogues.  Our aim is to explore a range of reasonable models for the galaxy--halo connection, and to understand the sensitivity of the cosmological signals to these models.

The paper is laid out as follows: In Sections \ref{sec:ds} and \ref{sec:hod_sham}, we describe the N-body simulations and the different galaxy models used to populate halos.
In Sections \ref{sec:rsd_linear} and \ref{sec:rsd_nonlinear}, we outline the theory describing two-point clustering statistics in redshift space and describe the dispersion model used in this paper. In Sections \ref{sec:populatinghalos}   and \ref{sec:rsd_nl}, we present our main results, showing the  redshift-space clustering signal from the different galaxy models and our approach for isolating nonlinear growth and redshift-space distortions in different $\mu$ bins. In Section \ref{sec:extract_b}, we demonstrate how the nonlinear bias can be extracted from the $\mu<0.2$ simulation data and carry out a joint parameter estimation to constrain the growth rate of structure in Section \ref{sec:params}.  We show that the growth rate can be extracted robustly for all of the galaxy models considered, to significantly smaller scales than is possible with current methods.

\section{Simulations \& galaxy population models
{\label{sec:galaxymodels}}}
In Section \ref{sec:ds} we describe the Dark Sky Gpc simulation used in this analysis. In Section \ref{sec:hod_sham} we outline the halo occupation distribution (HOD) model used as well as the different subhalo abundance matching methods employed to create mock galaxy samples.

\subsection{Dark Sky Simulations \label{sec:ds}}
Accurate RSD clustering measurements require both a large simulation volume, in order to recover linear theory predictions precisely on large scales, and high mass resolution, in order to resolve the velocity fields in the quasi-nonlinear regime. High mass and force resolution are also essential for accurately forecasting the nonlinear growth in both the velocity and density fields, as well as for resolving the halo substructure within virialized haloes needed to assign realistic galaxy populations.

The {\it Dark Sky} simulations\footnote{{\tt http://darksky.slac.stanford.edu}} are an unprecedented series of cosmological N-body simulations that evolve the large-scale structure of the Universe with high resolution over very large volumes \citep{2014arXiv1407.2600S}.  The two largest simulation boxes each followed the evolution of more than a trillion particles, over 1 and 8 \hGpc volumes, and were run using the 2HOT code \citep{2013arXiv1310.4502W} on the Titan machine at Oakridge National Laboratory.  For this study, we use the Dark Sky Gpc simulation ({\tt ds14\_b}), which follows the evolution of the matter distribution within a cubic region of 1 \hGpc on a side with 10240$^3$ particles, each with a particle mass of $m_p = 7.6\times 10^7 h^{-1}M_{\sun}$. 

These simulations adopt a $\Lambda$CDM cosmology which is compatible with Planck results \citep{2014A&A...571A...1P}. The cosmological parameters of the simulation are $\Omega_m = 0.295, \Omega_{\Lambda}=0.705, H_0 = 68.8$km s$^{-1}$Mpc$^{-1}$ and $\sigma_8 = 0.83$, where $\sigma_8$ is is the variance of the smoothed density field on a scale of 8Mpc$h^{-1}$ defined as \begin{eqnarray}
\sigma^2_8= \frac{1}{\left(2\pi^2\right)}\int_0^{\infty} {\rm d ln}k k^3 P(k) W^2(k,R=8) 
\end{eqnarray}
where $W(k,R=8)$ is the Fourier transform of a top hat window function.

Because of the large data volume for each full dark matter snapshot, each snapshot was downsampled with 1/32 of the total number of particles used in the calculation.  Full resolution snapshots were saved for a smaller number of timesteps, but in this work 
we use halo catalogs and merger trees constructed from 100 of these
downsampled snapshots. The halo and subhalo catalogues were made using the {\sc Rockstar} halo finder \citep{2013ApJ...762..109B} to locate gravitationally bound structures of 20 or more particles per halo. This halo-finding approach is based on adaptive hierarchical refinement of friends-of-friends groups in both position and velocity, using this phase space information to locate substructures and track subhalos better in the inner regions of halos.  Merger trees were created from these 100 downsampled snapshots, using the {\tt consistent-trees} code \citep{2013ApJ...763...18B}; these were used in creating the galaxy catalogs described in the following section. We focus our work on one snapshot of the simulation, at $z=0.67$.  Here, we can accurately measure the clustering signal in real and redshift space with nearly 100 million halos.

The power spectrum was computed by assigning the particles to a mesh using the cloud in cell (CIC) assignment scheme
and then performing a fast Fourier transform on the density field. Throughout this paper, the fractional error on the power spectrum plotted is given by $\sigma_P/P = (2/N)^{1/2}(1 + \sigma_n^2/P)$, where $N$ is the number of modes measured in a spherical shell of width $\delta k$  and $\sigma_n$ is the shot noise \citep{1994ApJ...426...23F}. This number depends upon the survey volume, $V$, as $N = V 4\pi k^2 \delta k/(2\pi)^3$.
Note that for this study the assumption of uncorrelated
errors is justified as it is expected that
each Fourier mode evolves independently \citep[see e.g. fig. 5 of ][]{2014MNRAS.443.1065B}.

\subsection{Galaxy samples
\label{sec:hod_sham}}
In this paper, we present redshift-space clustering measurements for several mock LRG samples at a redshift $z=0.67$ and number density $\bar{n} = 3.9 \times 10^{-4}$ (Mpc$/h)^{-3}$.  This corresponds to the number density of the SDSS-III {\tt cmass} sample at this redshift, and is roughly what is expected over 14,000 square degrees for the DESI survey \citep{2013arXiv1308.0847L}. However, we expect that the primary methodology we present here is also applicable to other sample definitions and to samples selected from other surveys.  We shall explore this further in future work.
 
Galaxies are biased tracers of the underlying dark matter distribution and are thought to reside in the potential wells (density peaks) of the dark matter field. Theoretical models for how galaxies occupy these dark matter halos are an essential step in connecting predictions from N-body simulations to galaxy surveys.  In the absence of full, hydrodynamic simulations which explicitly include the effects of star formation and feedback, populating a dark matter simulation with galaxies requires a detailed model to
connect the dark matter with the galaxies. In this paper, we consider two such modeling approaches: the halo occupation distribution (HOD) model and subhalo abundance matching models.  Our primary goal is to span a reasonable range of galaxy assignment schemes encompassing current theoretical uncertainty, and to understand the sensitivity of cosmological observables to these different assumptions.

%Here we focus on one type of galaxy sample, corresponding roughly to the luminous red galaxy (LRG) sample, with parameters similar to those 
%of the SDSS-III CMASS sample.  
%\risa{is it CMASS, or somehow DESI, with CMASS HOD?}
%that would be observed by the DESI survey at $z=0.67$.

The HOD model \citep[e.g.][]{2002ApJ...575..587B} describes the galaxy--halo connection by modeling the probability that a halo of  fixed virial mass $M$, hosts  $N$ galaxies, $P(N|M)$. The parameterization  of the HOD we use follows \citet{2005ApJ...633..791Z},  which separately models central and satellite galaxies, assuming that a central galaxy is required for a given halo to host a satellite. This model has been used in a number of studies; here, we use the best-fit parameters to this model from SDSS-III CMASS sample \citep{2014MNRAS.444..476R}, which is in basic agreement with previous HOD modeling of SDSS LRG samples \citep{2009ApJ...707..554Z, 2009ApJ...698..143R}.

In this HOD model, the probability for a halo of mass $M$ to host a central galaxy is
\begin{eqnarray}
N(M) &=& 0.5 \{  1+ {\rm erf}\left( \frac{ {\rm log}_{10}M   - {\rm log}_{10}M_{\rm min}}{\sigma_{{\rm log}_{10} M}}\right) \} \, .
\end{eqnarray}
The number of satellites assigned to the halo, given that it already hosts a central galaxy, is drawn from a Poisson distribution with mean
\begin{eqnarray}
N_{\rm sat} &=&  \left( \frac{M-M_{\rm cut}}{M_1}\right)^{\alpha} \,.
\end{eqnarray}
In the following we use ${\rm log}_{10}M_{\rm min} = 13.031\pm0.029$, $\sigma_{ {\rm log}{10} M}= 0.38\pm 0.06$, $ {\rm log}_{10} M_{\rm cut}=13.27\pm0.13$, $ {\rm log}_{10}M_1 = 14.08\pm 0.06$ and $\alpha = 0.76\pm 0.18$  \citep[as given by][]{2014MNRAS.444..476R}.
For both central and satellite galaxies, we assign velocities to each based on the centre-of-mass velocity of the subhalo found by the {\sc Rockstar} halo finder.  Satellite galaxies are assigned directly to subhalos identified by {\sc Rockstar}, ranked by their maximum circular velocity to match the number of satellites specified by the HOD. 

The semi-empirical approach of subhalo abundance matching \citep{2004ApJ...609...35K,2004MNRAS.353..189V} is based on the
assumption that some halo property is monotonically related
to some galaxy property, typically galaxy luminosity or stellar
mass. A natural first assumption is that galaxy properties are strongly correlated with the depth of their potential wells and in this case the maximum circular velocity of a halo (or subhalo) at the present time, {\em \vmax}, would be the relevant property.  Given that dark matter halos can be significantly stripped, either before or after they enter the virial radius, in a way that galaxies are not, several authors have shown that models that instead associate galaxy properties with subhalos before they start getting stripped (for example at accretion on to the main halo, or at the maximum mass they had in their accretion history) provide a better match to data \citep[e.g.][]{2006ApJ...647..201C}.

\citet{2013ApJ...771...30R} carried out a detailed study of the underlying assumptions of the subhalo abundance matching technique including which halo property is most closely associated with galaxy stellar masses and luminosities, and how much scatter is in this relationship. These authors find that an abundance matching model that associates galaxies with maximum value of the  halo maximum circular velocity \vmax\ that a halo had over its accretion history (\vpeak), is in good agreement with the data, when scatter of 0.20 $\pm $0.03 dex in stellar mass at a given \vpeak\ is included.
Subsequently, \cite{2014ApJ...787..156B} pointed out that this peak circular velocity over a halo's history is generally determined by the time of the last major mergers, which may make it less physically motivated as the most appropriate abundance matching proxy.  \citet{2015arXiv151005651L} suggest instead to consider the maximum circular velocity at the peak value of the {mass} over its history, \vmpeak, which is less impacted by mergers (a similar motivation was used by \citealt{2015arXiv150701948C} to suggest the proxy they refer to as $v_{\rm relax}$).  This work further discusses a set of possible models that vary the impact of assembly bias \citep{2006ApJ...652...71W},
ranging from the maximum mass that the halo (or subhalo) has ever had in its merger history,
{\em \mpeak} and the maximum velocity at the time when the halo (or subhalo) has achieved the maximum mass in its merger history, {\em \vmpeak}.

In this work, our goal is to span a reasonable range of galaxy assignment schemes encompassing current theoretical uncertainty.  Thus in addition to the HOD models, we consider three representative subhalo abundance matching models, using the proxies \mpeak, \vmpeak, and \vmax\ to rank order halos. All three models predict different satellite fractions and clustering amplitudes in real space, and have differing impact of assembly bias, in contrast to the HOD model.  In this work we present, for the first time, predictions for the clustering signal in redshift space for all three models.  This allows us to investigate the sensitivity of the redshift-space clustering to these differences.
For ease of comparison the number density for each is matched to that from the HOD. For the  \vmpeak\ catalog this corresponds to a magnitude cut of $M_r < - 21.36$. 

%Many ways exist to abundance match observed galaxies to
%dark matter halos in simulations and the three proxies we have chosen reasonably span the model space \citep[see e.g.][]{2013ApJ...770...57B,2013ApJ...771...30R,Lehmann2015}. E.g. the mpeak proxy has been used successfully to reproduce  galaxy conditional stellar mass functions as well as galaxy clustering as a function of stellar mass or luminosity
%and redshift \citep[][]{2006ApJ...647..201C,2013ApJ...771...30R,2013ApJ...772..139W}.
%All three models predict different satellite fractions and clustering amplitudes in real space. 

In modeling the bias of these samples, we choose the simple Q-model for nonlinear bias \citep{2005MNRAS.362..505C}
 \begin{eqnarray}
 b_{\rm nl} &=& b_{\rm lin}\sqrt{\frac{1+Qk^2}{1+Ak}},
 \label{eq:nlbias}
 \end{eqnarray}
 where the variables $b_{\rm lin}$ and $Q$ are allowed to vary while $A$ is kept fixed at a value 1.7 \citep[see][]{2005MNRAS.362..505C}.

\section{Redshift-space distortions 
{\label{sec:rsd_theory}}
}
In Section \ref{sec:rsd_linear}, we discuss the linear perturbation theory predictions for  two-point clustering statistics in redshift space. In Section \ref{sec:rsd_nonlinear}, we outline the simple dispersion model we shall use in this paper.

\subsection{Linear perturbation theory \label{sec:rsd_linear}}

Inhomogeneous structure in the Universe induces peculiar motions which distort the clustering
pattern measured in redshift space on all scales. This effect must be taken into account when analyzing three-dimensional datasets that use
redshift to estimate the radial coordinate.
Redshift-space effects alter the appearance of the clustering
of matter, and together with nonlinear evolution and bias, lead the measured
power spectrum to depart from  the simple predictions of linear perturbation theory.
The comoving distance to a galaxy, $\vec{s}$,  differs from its true distance, $\vec{x}$, due to its peculiar velocity, $\vec{v}(\vec{x})$
(i.e. an additional velocity to the Hubble flow). The mapping from redshift space to real space is given by
\begin{equation}
\vec{s} = \vec{x} + u_z \hat{z},
\end{equation}
where $u_z = \vec{v}\cdot \hat{z}/(aH)$ and $H(a)$ is the Hubble parameter. This assumes that the distortions take place along the line of sight, denoted by $\hat{z}$, and is commonly referred to as the plane--parallel approximation.

On small scales, randomized velocities associated with the motion of galaxies inside virialized structures reduce the power.
The dense central regions of galaxy clusters appear elongated along the line of sight in redshift space, which produces the  \lq fingers
of God\rq\
effect in redshift survey plots.
For growing perturbations on large scales, the overall effect of redshift-space distortions is to enhance the clustering amplitude.
Any difference in the velocity field due to mass flowing from underdense regions to high density regions will alter the volume element, causing
an enhancement of the apparent density contrast in redshift space, $\delta_s(\vec{k})$, compared to that in real space, $\delta_r(\vec{k})$ \citep[see][for a review of RSDs]{1998ASSL..231..185H}.

Assuming the line of sight component of the peculiar velocity
is along the $z$-axis, the power spectrum in redshift space is given by \citep[see e.g.][]{Scoccimarro:2004tg}
\begin{align}
\delta_D(\vec{k}) + P_s(\vec{k}) =\int \!\! \frac{{\rm d}^3r}{(2\pi)^3} e^{-i\vec{k}\cdot\vec{r}} \langle e^{ik_zV}[1+\delta_g(\vec{x})][1+\delta_g(\vec{x}')]\rangle
\end{align}
where $\delta_g = b \delta$ is the galaxy overdensity which is usually related by a linear bias, $b$ to the matter overdensity, $V = u_z(\vec{x}) - u_z(\vec{x}')$ and $\vec{r} = \vec{x} - \vec{x'}$. We are also assuming that there is no velocity bias between the dark matter and galaxies for simplicity, although this assumption may not be true in detail 
\citep[e.g.][]{2014PhRvD..90j3529B,2014arXiv1410.1256Z,2015MNRAS.446..793J}

Decomposing the vector field into curl- and divergence-free parts, and assuming an irrotational velocity field, we can re-write
$k_z u_z =  -(k_z^2/k^2 )\theta(k)  = -\mu^2 \theta(k)$ where $\theta(k)$ is the Fourier transform of the velocity divergence.
Expanding the exponential term and only keeping terms up to second order in the variables $\delta$ and $\theta$, the power spectrum in redshift space $P_s$ becomes
\begin{align}\label{eq:nl_rsd}
\delta_D(\vec{k} - \vec{k'})P_s(\vec{k}) &=b^2\langle \delta(\vec{k})\delta^*(\vec{k'}) \rangle -2\mu^2 b \langle \theta(\vec{k})\delta^*(\vec{k}') \rangle \nonumber \\
&+ \mu^4\langle \theta(\vec{k})\theta^*(\vec{k}') \rangle.
\end{align}
If we assume the linear continuity equation holds we can re-write this as
\begin{align}
\nonumber
\delta_D(\vec{k} - \vec{k'})P_s(\vec{k}) &=\langle \delta(\vec{k})\delta^*(\vec{k'})\rangle [b^2  -2bf\mu^2 + f^2 \mu^4] \\ \nonumber
&=  \delta_D(\vec{k}-\vec{k'})P(k)[b^2  - 2bf\mu^2 + f^2 \mu^4]  \label{eq:kaiser}\\
\end{align}
which is the \citet{1987MNRAS.227....1K} formula for the power spectrum in redshift space in terms of the linear bias $b$, the power spectrum $P(k)$
and the linear growth rate $ f$, given by
\begin{eqnarray}
f =  \frac{{\rm{d ln}}D}{{\rm d ln}a} \, ,
\end{eqnarray}
where $D$ is the linear growth factor.

Rather than using the full 2D power spectrum, $P(k,\mu)$, it is common to decompose
the matter power spectrum in redshift space into multipole moments using Legendre polynomials, $L_l(\mu)$, \citep[see e.g.][]{1998ASSL..231..185H}
\begin{eqnarray}
P(k,\mu) = \sum_{l} P_l(k) L_l(\mu) \, ,
\end{eqnarray}
where the summation is over the order, $l$, of the multipole.
The anisotropy in $P( \vec{k} )$ is symmetric in $\mu$, as $P(k,\mu)=P(k,-\mu)$, so only even values of $l$ are summed over. Each multipole moment is given by
\begin{eqnarray}
P^s_l(k) = \frac{2l+1}{2} \int_{-1}^{1} P(k,\mu) L_l(\mu) \rm{d}\mu \, ,
\end{eqnarray}
where the first two non-zero moments have Legendre polynomials, $L_0(\mu) = 1$ and  $L_2(\mu) = (3\mu^2 - 1)/2$.
Using the linear model in Eq. \ref{eq:kaiser}, the first two multipole moments are given by
\begin{eqnarray}
\left( {\begin{array}{c}
 P_0(k)   \\
 P_2(k) 
 \end{array} } \right)
 &=& P_{dm}(k)
\left( {\begin{array}{c}
1 + \frac{2}{3}\beta + \frac{1}{5}\beta^2   \\
\frac{4}{3}\beta + \frac{4}{7}\beta^2 
 \end{array} } \right) \, ,
\label{moments1}
\end{eqnarray}
where $P_{dm}(k)$ denotes the real space matter power spectrum. Note we have 
omitted the superscript $s$ here
for clarity. The variable $\beta = f/b$ is the ratio of the linear growth rate to the bias.

\subsection{Modeling $P(k,\mu)$ in the nonlinear regime \label{sec:rsd_nonlinear}}

The equations in Section \ref{sec:rsd_linear} describe the boost in the clustering signal in redshift space on large scales where linear perturbation theory is valid.
To go beyond linear theory and deal with small-scale velocities requires a model for the velocity field and all the density velocity correlations.

Commonly used models for the redshift-space power spectrum extend the Kaiser formula by assuming that the velocity and density fields are uncorrelated and that the joint
probability distribution factorizes as  $\mathcal{P}(\delta,\theta) = \mathcal{P}(\theta)\mathcal{P}(\delta)$.
Examples include multiplying Eq. (\ref{eq:kaiser}) by a factor which
attempts to take into account small-scale effects, invoking either a
Gaussian or exponential distribution
of peculiar velocities.
A popular phenomenological example of this which incorporates the damping effect of velocity dispersion on small scales is the so-called \lq dispersion model\rq \ \citep{1994MNRAS.267.1020P},
\begin{equation}
P^s(k,\mu) =  P_g(k) (1+\beta \mu^2)^2 \frac{1}{(1 + k^2 \mu^2 \sigma_p^2/2)} 
\label{eq:pd}\, ,
\end{equation}
where $P_g$ is the galaxy power spectrum, $\sigma_p$ is the pairwise velocity dispersion along the line of sight, which is treated as a parameter to be fitted to the data. In this paper we model the damping effect using an exponential term as 
\begin{eqnarray}
P^s(k,\mu) =  P_g(k) (1+\beta \mu^2)^2 e^{ -(k \mu \sigma_v)^2} \, . 
\label{eq:exp}
\end{eqnarray}
This model has been used to fit to results from both simulations and observations 
\citep[see, for example][]{Scoccimarro:2004tg,2009MNRAS.393..297P,2011MNRAS.410.2081J,2008Natur.451..541G, 2011MNRAS.415.2876B, 2012MNRAS.423.3430B}.  The dispersion model is a simplification in which not only are density--velocity correlations neglected, the velocity fields are assumed to be linear and the velocity dispersion is scale independent. Each of these assumptions could impact the accuracy of the linear growth rate extracted from two-point clustering statistics \citep[see e.g.][]{2015MNRAS.449.3407J}.

Recently, many models have been presented that improve on this description of redshift-space distortions in the nonlinear regime \citep[e.g.][]{Scoccimarro:2004tg, 2008PhRvD..78h3519M, 2010PhRvD..82f3522T, 2011MNRAS.417.1913R, 2011JCAP...11..039S}.
Many of these models still require a free parameter to fully describe the velocity dispersion effects and can only accurately recover the linear growth rate on surprisingly large scales, e.g.  $k<0.2h/$Mpc$^{-1}$ \citep[][]{2011MNRAS.410.2081J, 2012ApJ...748...78K, 2015MNRAS.447..234W}.
In all models for the power spectrum in redshift space, there is a degeneracy between the galaxy bias, the linear growth rate, and the amplitude of fluctuations $\sigma_8$ \citep{2009MNRAS.393..297P}. Introducing a free parameter to describe the
scale-dependent damping due to velocity distortions on small scales adds to this degeneracy, and inevitably weakens constraints on the growth rate.
In this paper we examine the full 2D power spectrum $P(k,\mu)$, rather than using the multipoles as in previous studies. This allows us to examine the impact of RSD and nonlinear growth separately and use appropriate models to extract the maximum information on both the linear growth rate and the bias along different angles with respect to the line of sight.

\section{Results}
In Section \ref{sec:populatinghalos}, we present the measured power spectrum in redshift space at $z=0.67$ using different models for populating the dark matter halos with galaxies. In Section \ref{sec:rsd_nl}, we demonstrate how it is possible to distinguish between the effects of nonlinear growth and redshift-space distortions along different angles with respect to the line of sight by examining the full 2D $P(k,\mu)$ in redshift space.
In Section \ref{sec:extract_b}, we examine the $\mu<0.2$ simulation data in detail in order to measure the nonlinear bias as a function of wavenumber.
We carry out a joint likelihood parameter estimation in Section \ref{sec:params}, in order to constrain both the linear growth rate, $\sigma_8$, and the linear bias.

\begin{figure*}
\begin{center}
\includegraphics[height=3in,width=7.in]{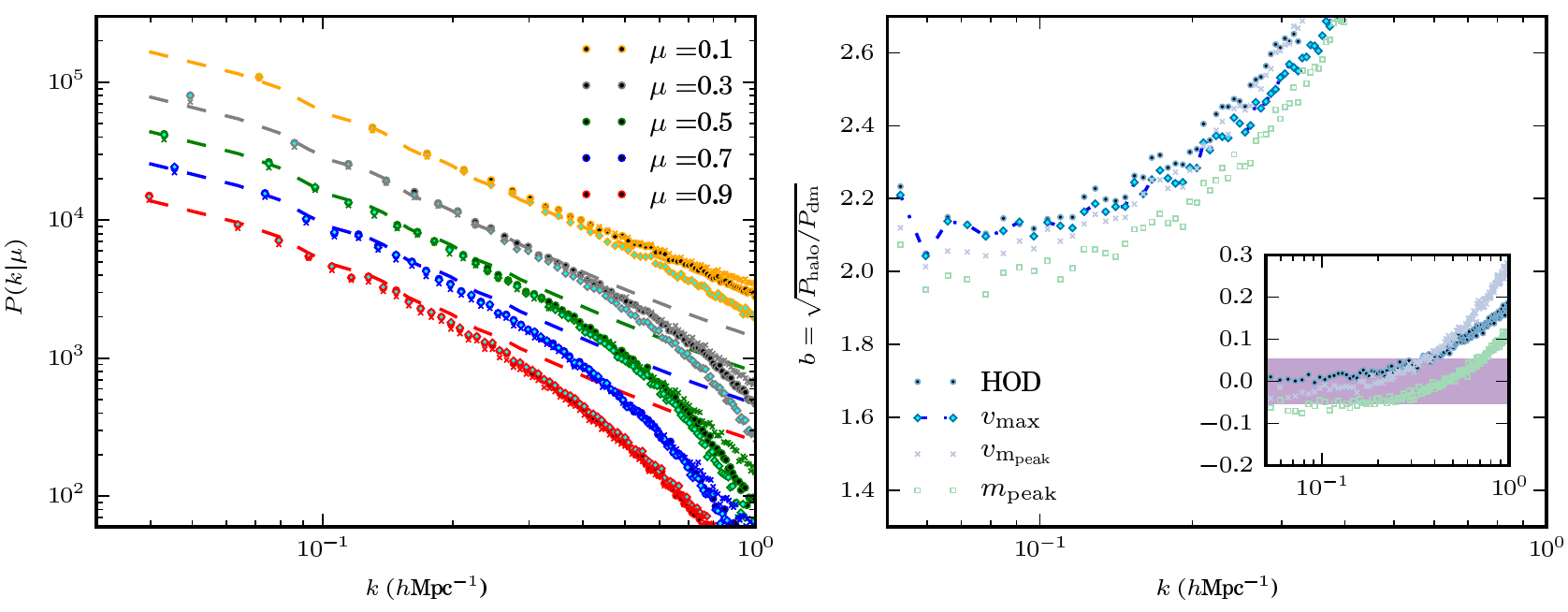}
\caption{{\em Left:} The halo $z=0.67$ anisotropic power spectra $P(k,\mu)$ in redshift space for the HOD (circles), \vmax (diamonds)  and \vmpeak (crosses) models.
The binned power spectra are shown as  $0\leq \mu<0.2$ (orange),  $0.2\leq \mu<0.4$ (grey),  $0.4\leq \mu<0.6$ (green),  $0.6\leq \mu<0.8$ (blue) and  $0.8\leq \mu<1$ (red).
The Kaiser model prediction using a nonlinear dark matter power spectrum and the best-fit linear bias for each has been plotted as a dashed line.
Each $\mu = 0.3 \,(0.1)$ and $\mu = 0.7 \,(0.9)$  power spectrum has been separated by a factor or 2 (2.5) from the $\mu=0.5$ data for clarity. 
{\em Right:} The nonlinear bias $b = \sqrt{P_H(k)/P_{\rm DM}}$ as a function of scale. The inset panel shows the ratio of the bias for \vmpeak (crosses), \mpeak (squares) and HOD (circles) to the nonlinear bias for \vmax as a function of scale. The shaded band shows $\pm 5$\% difference in the ratio.
}
\label{fig:pk_k_mu}
\end{center}
\end{figure*}

\subsection{Populating halos with galaxies and the effects on the RSD signal {\label{sec:populatinghalos}}}

In Fig. \ref{fig:pk_k_mu}, 
we plot the full 2D anisotropic power spectra in redshift space, $P(k,\mu)$, at $z=0.67$ as a function of wavenumber, $k$, for our simulated LRG sample. We have binned the measured simulation data into five bins, plotted in the left panel as $\mu=0.1$ (orange), $\mu=0.3$ (grey), $\mu=0.5$ (green), $\mu=0.7$ (blue), and $\mu=0.9$ (red) for four different methods of populating halos with galaxies.
Note that the bins in the power spectra have been offset from each other for clarity.   In Fig. \ref{fig:pk_k_mu} the HOD model is plotted as circles and the SHAM models: \vmpeak, \mpeak, and \vmax, are plotted as crosses, squares, and diamonds respectively. 
We plot the Kaiser model prediction using a nonlinear dark matter power spectrum and the best-fit linear bias for each $\mu$ bin as a dashed line.

From the left panel in Fig. \ref{fig:pk_k_mu}, it is clear that the different models for populating halos have differing redshift-space clustering on quasi-linear to nonlinear scales, $k> 0.1h$Mpc$^{-1}$, and that the measured clustering amplitudes differ from the Kaiser model, which includes nonlinear growth in the dark matter field. Only the $\mu=0.1$ bin (orange points) seems to agree with the Kaiser model plotted, however there are still significant differences between all SHAM models and the HOD on scales $k> 0.4h$Mpc$^{-1}$.

In linear perturbation theory $P(k,\mu) = (b + f\mu^2)^2P_r(k)$  and so we would expect the $\mu = 0.1$ bin (orange) to be least affected  by redshift-space distortions while the $\mu = 0.9$ bin (red) should be most impacted. By comparing each to a linear theory Kaiser model with a nonlinear dark matter power spectrum, as in Fig. \ref{fig:pk_k_mu}, we can get an approximate sense of the relative impact of {\em nonlinear} RSD effects, as distinct from the usual nonlinear growth in real space and the boost in clustering on large scales due to coherent flows, on each $\mu$ bin. We find that the impact of nonlinear RSD also scales 
with $\mu$, as expected from the nonlinear damping model in Eqn. \ref{eq:exp}. From this figure we can see there is little damping of the $\mu=0.1$ bin due to nonlinear RSD while the $\mu=0.9$ bin is most impacted.

In the right panel of Fig. \ref{fig:pk_k_mu}, we plot the  nonlinear bias $b = \sqrt{P_H(k)/P_{\rm DM}}$ as a function of scale for each model, where 
$P_H(k)$ is the real space halo power spectrum and $P_{\rm DM}$ is the linear matter power spectrum. Here the HOD, \vmpeak, \mpeak and \vmax models are
plotted as circles, crosses, squares and diamonds respectively.  Only the HOD and \vmax models have similar large scale bias.
The inset panel shows the ratio of the bias for each model to the bias for the \vmax  model as a function of scale. The shaded band shows $\pm 5$\% difference in the ratio and it is clear that on scales $k\ge0.4h$Mpc$^{-1}$ the difference between the models is about 10\%.

\begin{figure}
\begin{center}
\includegraphics[height=2.8in,width=3.in]{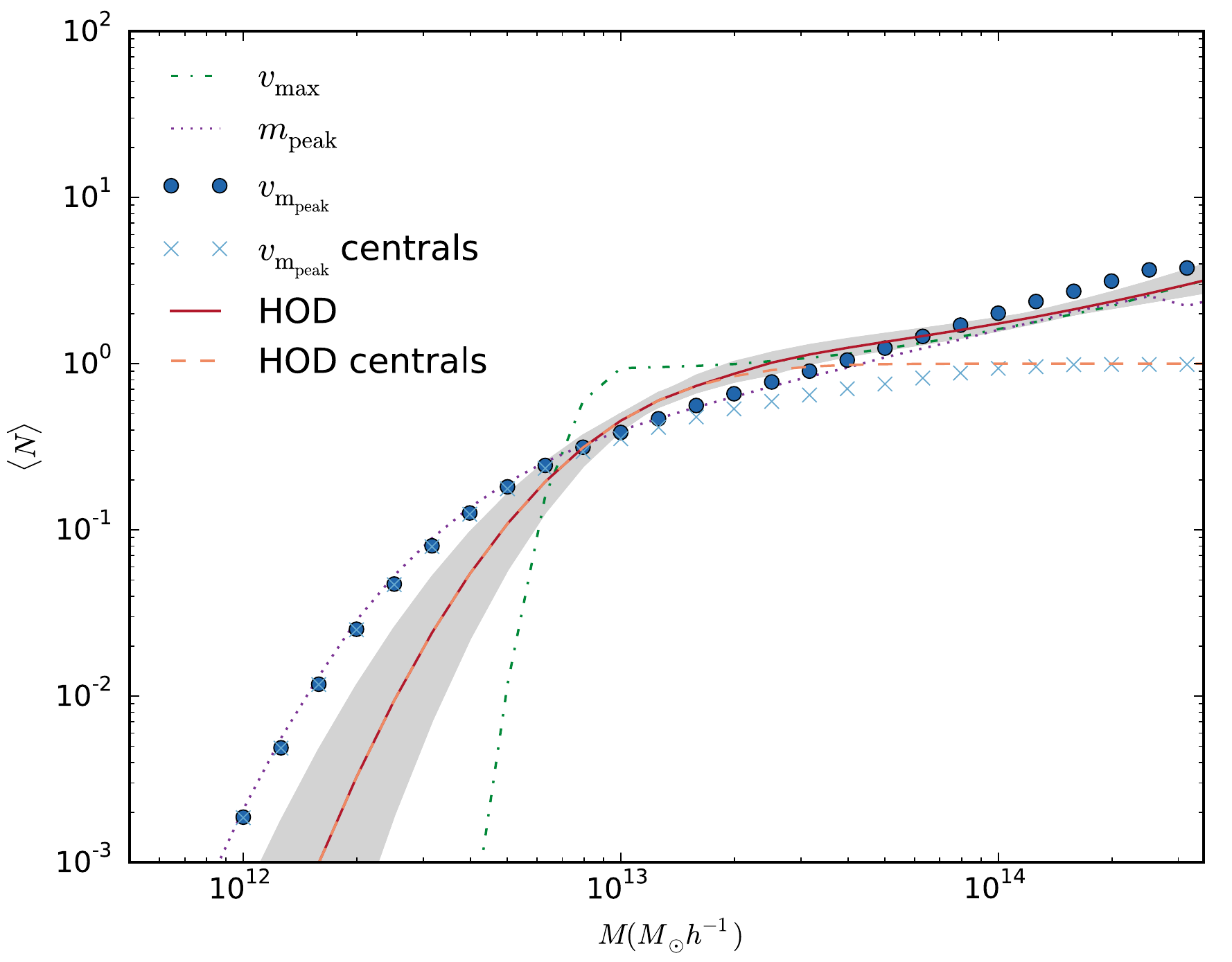}
\caption{
 The HOD model (solid red line) at $z=0.67$. The contribution from centrals only is shown as an orange dashed line. The measured number density from the \vmpeak, \mpeak, and \vmax\ catalogues are shown as blue circles, dotted and dot-dashed lines respectively. The \vmpeak\ centrals only are shown as light blue crosses. The shaded grey region represents 1500 random samples from the $1-\sigma$ parameter range of the HOD.
}
\label{fig:hod}
\end{center}
\end{figure}

These differences in the nonlinear bias arise from differences in both the central and satellite populations for the different models. 
In Fig. \ref{fig:hod}, we show the HOD used in this paper as a solid red line (the contribution from centrals only is shown as an orange dashed line). The shaded grey region represents 1500 random samples from the $1-\sigma$ parameter range of the HOD. The measured HOD from the \vmpeak, \mpeak\ and \vmax\ catalogues are shown as blue circles, dotted and dot-dashed lines, respectively. The light blue crosses represent $\langle N \rangle$ for \vmpeak\ centrals only. 
The fact that these models are selecting different central and satellite populations  
gives rise to differences in both RSD and nonlinear growth of matter.
 With a fixed number density for each model the central fraction is highest for \vmax\ and is smaller for the HOD, \mpeak\ and \vmpeak\ (in decreasing order) as shown in Fig. \ref{fig:hod}.

 In the standard halo model \citep[see e.g.][]{1991ApJ...381..349S,2000ApJ...543..503M,2002ApJ...575..587B,2002PhR...372....1C}, which is a convenient formalism for predicting and interpreting the clustering statistics of dark matter halos and galaxies, the clustering signal can be written as a sum of one and two halo terms.
The one-halo term, due to distinct mass elements that lie within
the same dark matter halo, dominates the clustering signal on
scales smaller than the virial radii of halos and the two-halo term, which is due to mass elements in
distinct pairs of halos, dominates on scales
much larger than the virial radii of the largest halos.
In this context larger satellite fractions increase the one-halo term, boosting the nonlinear clustering signal and increasing  
velocity dispersions on small scales.
If we can attribute differences in the $\mu=0.1$ bin only to nonlinear growth in the matter field without RSD and consider differences in the
$\mu=0.9$ bin as due to a mix of RSD and nonlinear growth then, from
the left panel in Fig. \ref{fig:pk_k_mu}, we would anticipate that the \vmpeak sample has the largest satellite fraction, followed by the HOD and then
\vmax. Increased small-scale velocity dispersion is manifested as increased RSD damping in the clustering on small scales, clearly seen in the $\mu = 0.9$ (red) bin.
This interpretation seems valid as we find that $f_{\rm sat} = 0.151, 0.087$ and  $0.061$ for \vmpeak, the HOD
and \vmax\ models respectively. Note \mpeak\ is not plotted in the left panel for clarity ($f_{\rm sat} = 0.117 $ for \mpeak).

\begin{figure}
\begin{center}
\includegraphics[height=3.in,width=3.3in]{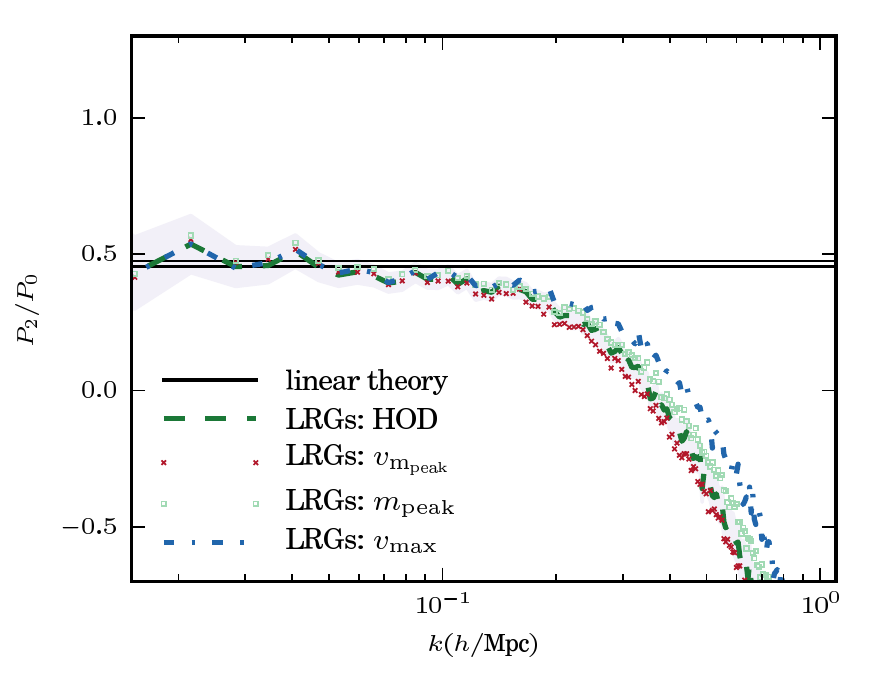}
\caption{The quadrupole to monopole ratio as a function of wavenumber at $z=0.67$ for the HOD (green dashed line), \vmax\ (blue dot dashed line), \mpeak\ (green squares)  and \vmpeak\ (red dotted) models.  The solid horizontal lines represent the linear theory predictions. Note there are two linear theory lines plotted due to different linear bias factors for the models.}
\label{fig:p2p0}
\end{center}
\end{figure}

In Fig. \ref{fig:p2p0}, we show the ratio of the 
quadrupole to monopole moment of the power spectra in redshift space  for  the HOD (dashed lines), \vmax\ (dot dashed lines), \mpeak\ (green squares)  and \vmpeak\ (dotted lines) models. The different linear bias between the models gives rise to linear theory predictions which are slightly different (thin and thick solid black lines).  We find that differences between the HOD, \vmax, \mpeak, and \vmpeak models are largest for the quadrupole $P_2$ moment on quasi-linear to nonlinear scales $k> 0.1h$Mpc$^{-1}$. This result agrees with previous findings that the quadrupole is most
 sensitive to differences in velocity dispersion and hence
 differences in the satellite fractions \citep{2014MNRAS.444..476R}. Many previous analyses of RSD have focused 
 on these multipole moments which integrate over the $\mu$ dependence. From Fig. \ref{fig:p2p0}
 we would correctly predict that the  \vmpeak sample has the largest satellite fraction, followed by the HOD and then
\vmax based on the degree of damping on scales  $k> 0.2 h$Mpc$^{-1}$. 
However the information content in individual $\mu $ bins seems richer, thus motivated by Fig. \ref{fig:pk_k_mu}, in the next section we shall investigate the idea of separating the nonlinear growth and RSD effects further. Although RSD effects on small scales are also the result of nonlinear growth, the distinction we make in this paper is between nonlinear growth in real space and the distortions along the line-of-sight due to peculiar velocities.
We shall analyse the differences in the nonlinear bias on small scales further in Section \ref{sec:extract_b}.

\subsection{Separating RSD and NL growth \label{sec:rsd_nl}}
 
\begin{figure*}
\begin{center}
\includegraphics[height=2.8in,width=3.in]{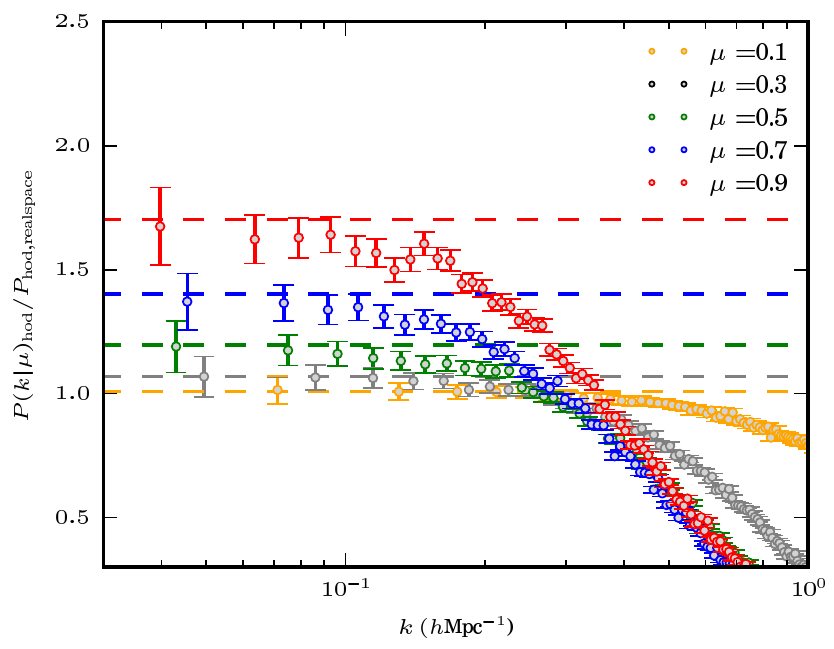}
\includegraphics[height=2.8in,width=3.in]{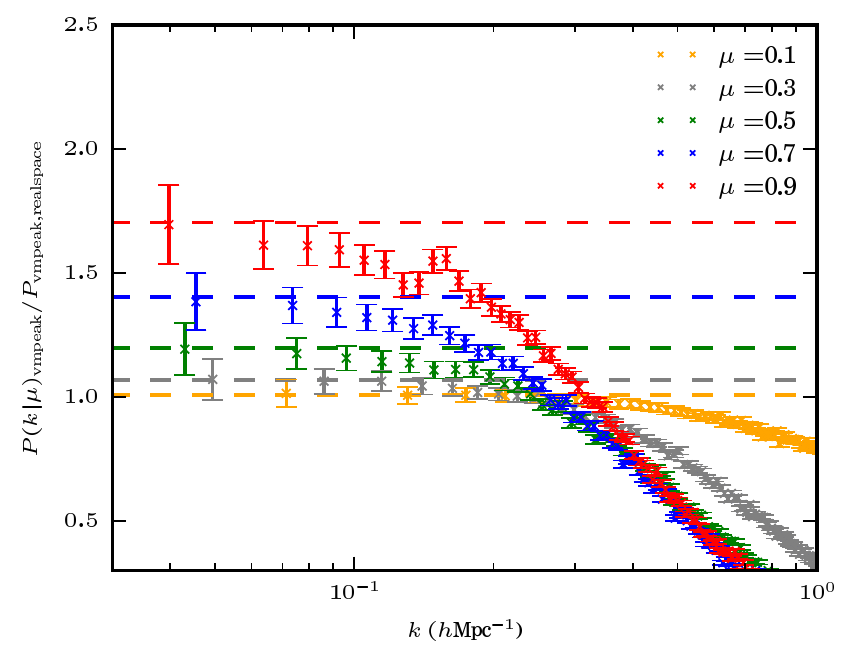}
\includegraphics[height=2.8in,width=3.in]{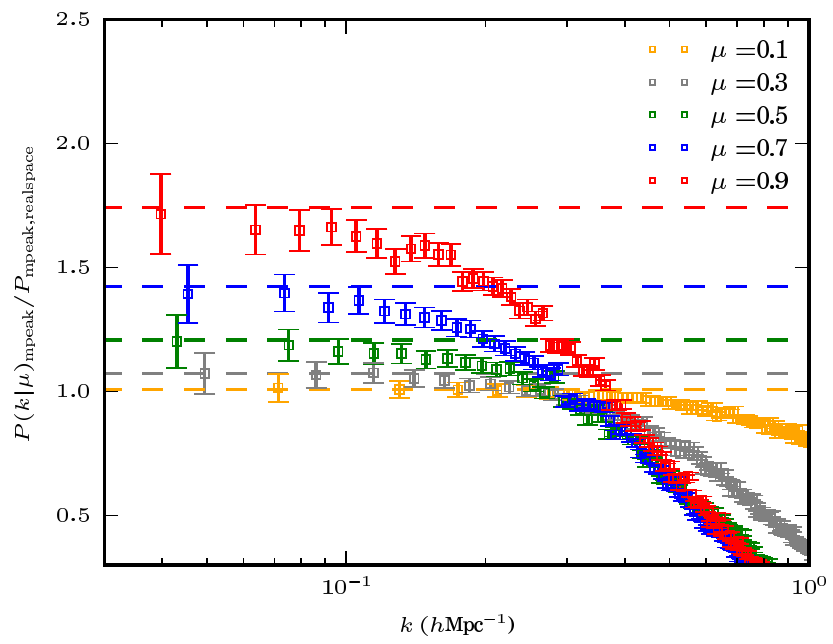}
\includegraphics[height=2.8in,width=3.in]{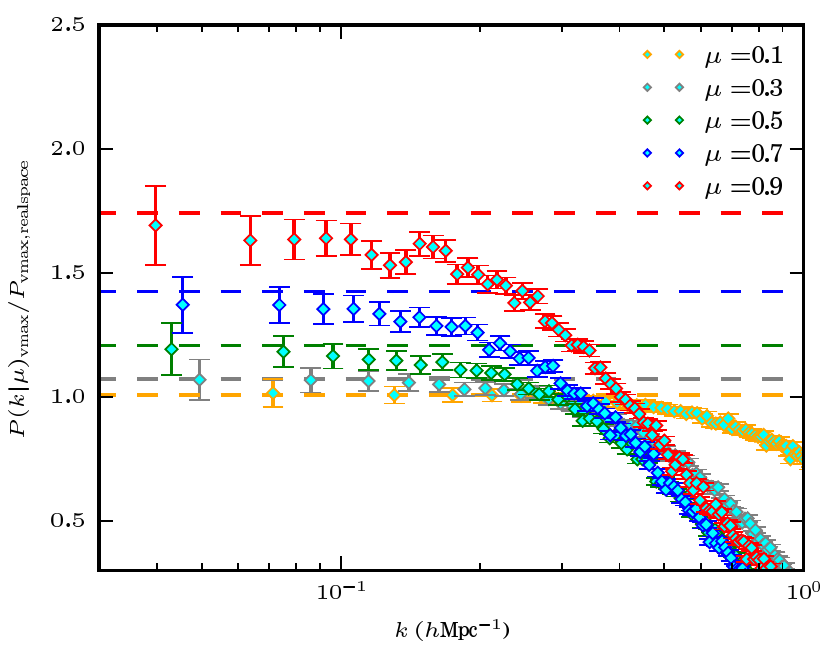}
\caption{
The ratio of the nonlinear power spectra in redshift space to real space,  $P(k,\mu)/P_r(k)$, for 4 different galaxy occupation models.  The colors of each line represent the $\mu$ bin plotted as given in the legend. The dashed lines correspond to the linear theory prediction $(1+f/b_{\rm lin} \mu^2)^2$ for each $\mu$ bin using the best-fit linear bias to each bin.
Panels correspond to the LRG HOD model (top left), the  \vmpeak\ sample (top right), the \mpeak\ sample (bottom left), and the \vmax\ sample (bottom right).
}
\label{fig:phod}
\end{center}
\end{figure*}

In the upper left (right) panel of Fig. \ref{fig:phod}, we plot the ratio of the nonlinear redshift-space power spectra, $P(k,\mu)$, to the nonlinear real-space power spectrum  measured from the simulations using the HOD (\vmpeak) model. The dashed lines correspond to the linear theory prediction $(1+f/b_{\rm lin} \mu^2)^2$ for each $\mu$ bin
using the best-fit linear bias for each model. By dividing by the nonlinear matter power spectrum, we are isolating the RSD effects. 
The ratio for each $\mu$ bin is not unity on large scales, due to the boost in the clustering signal caused by coherent bulk flows, and agrees with linear theory predictions on different scales depending on the bin. E.g. the $\mu=0.5$ data agrees with linear theory at $k<0.15 h$Mpc$^{-1}$ whereas the $\mu=0.9$ data
agrees with linear theory on scales $k<0.06 h$Mpc$^{-1}$.
In the lower left (right) panel of Fig. \ref{fig:phod}, we show similar ratios for the \mpeak\ and \vmax\ models respectively.
What is striking from all panels in Fig. \ref{fig:phod} is that the ratio for the orange $\mu=0.1$ bin is unity on scales $k<0.4 h$Mpc$^{-1}$. This means that on these scales there are negligible redshift-space distortion effects for the $\mu=0.1$ bin.
Note also that there is a increase in the
nonlinear RSD effects which impacts larger and larger scales as $\mu$ increases.

\begin{figure}
\begin{center}
\includegraphics[height=2.8in,width=3.in]{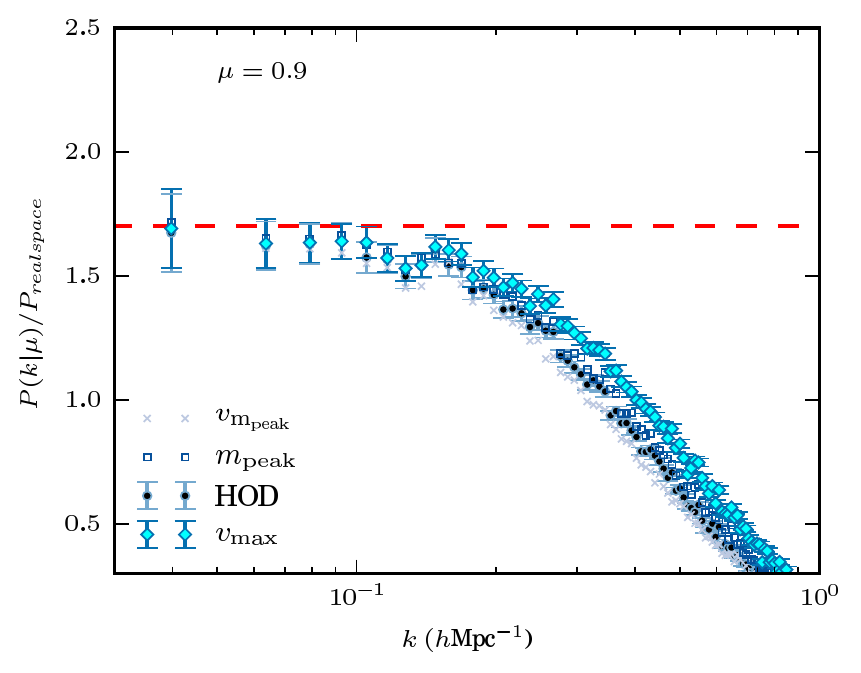}
\caption{The ratio of the nonlinear power spectra in redshift space to real space,  $P(k,\mu)/P_r(k)$  for all models, HOD (circles), \mpeak\ (squares), \vmpeak\ (crosses) and \vmax\ (diamonds) in the $\mu=0.9$ bin. We do not plot error bars for \mpeak\ or \vmpeak\ for clarity. The linear theory prediction for the HOD model is shown as a dashed red line. 
}
\label{fig:mu09}
\end{center}
\end{figure}

In order to highlight the differences between the four models we plot the $\mu=0.9$ data only in Fig. \ref{fig:mu09} for the 
 HOD (circles), \mpeak\ (squares), \vmpeak\ (crosses) and \vmax\ (diamonds) model. The linear theory prediction for the HOD model is shown as a dashed red line. This plot highlights the differences in RSD effects between the models and agrees with our findings in the previous section, that models with higher satellite fractions like \vmpeak\ have higher FOG damping signal on quasi-linear to nonlinear scales compared to models with a lower satellite fraction like \vmax. At a scale of $k=0.3 h$Mpc$^{-1}$, there is approximately a 25\% difference between the \vmax\ and \vmpeak\ model
  in the damping signal in the $\mu=0.9$ bin.

\begin{figure*}
\begin{center}
\includegraphics[height=3.in,width=3.2in]{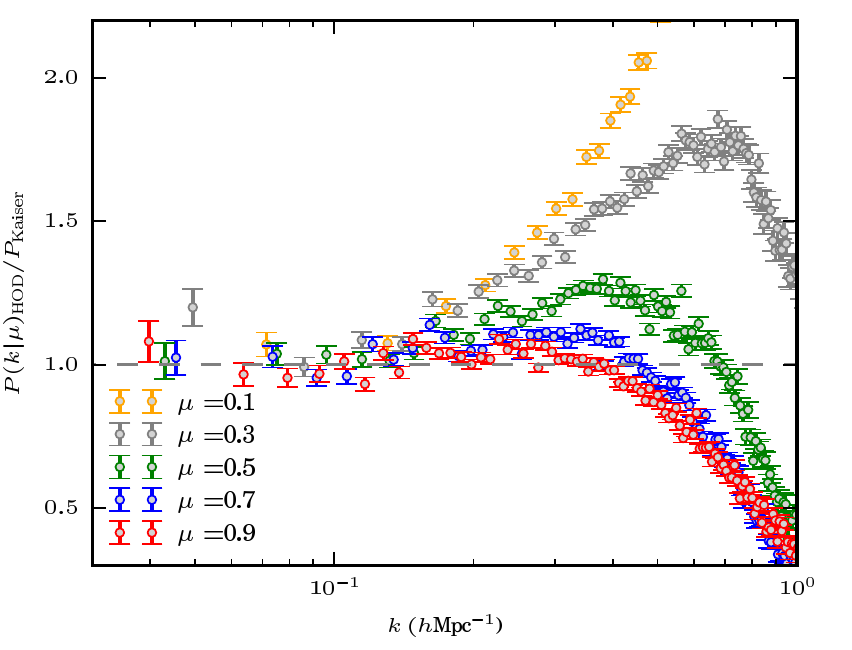}
\includegraphics[height=3.in,width=3.2in]{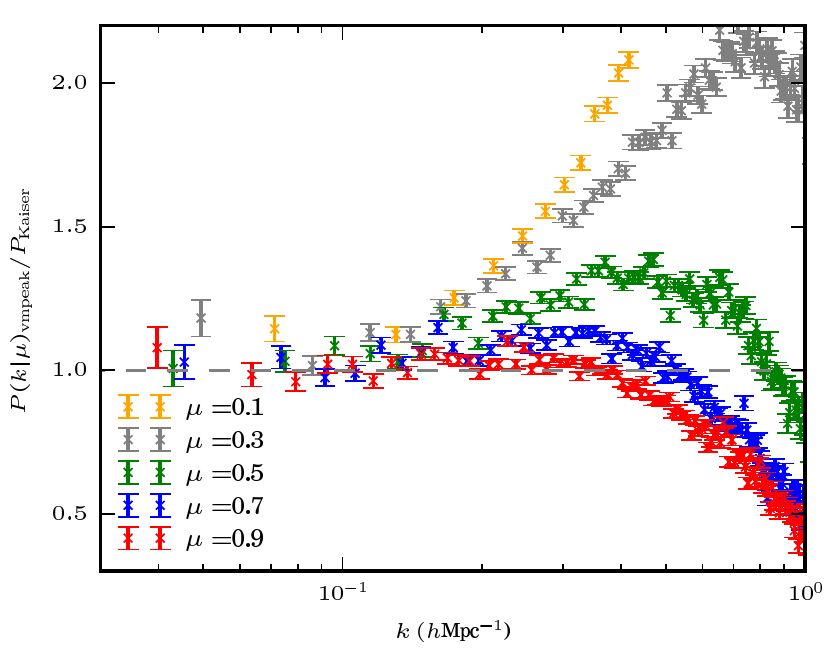}
\includegraphics[height=3.in,width=3.2in]{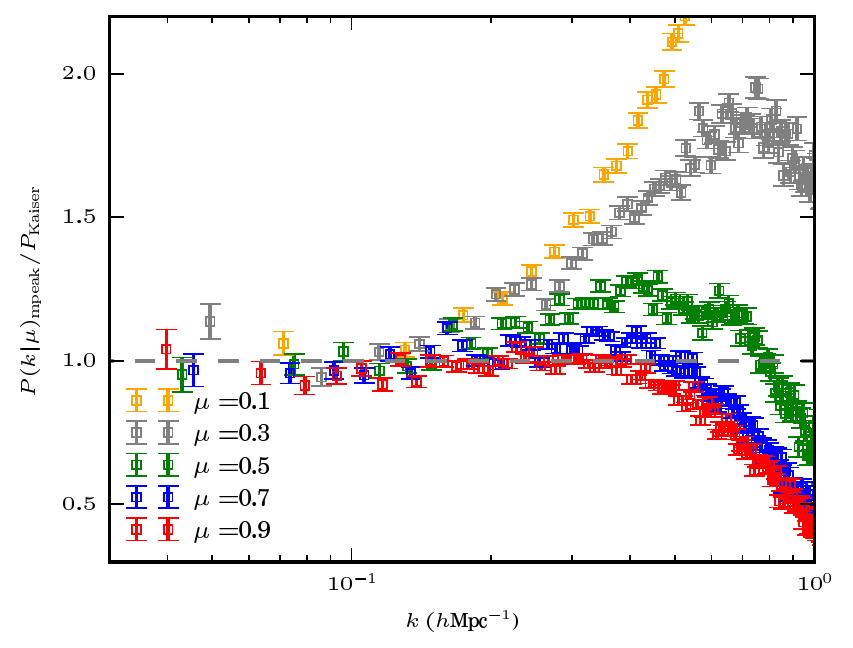}
\includegraphics[height=3.in,width=3.2in]{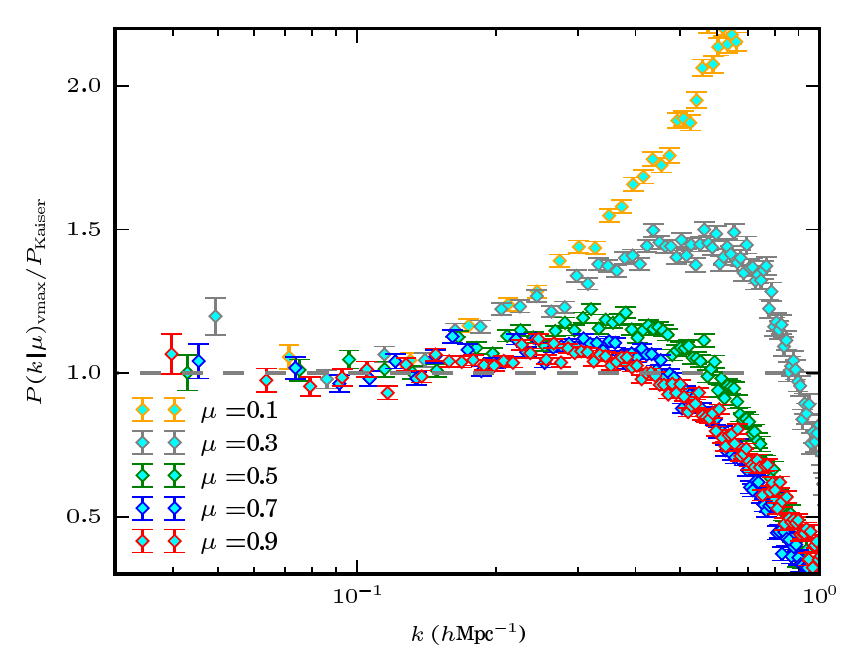}
\caption{
The ratio of the anisotropic redshift space power spectrum $P(k,\mu)$ for several models, 
compared to the  linear theory Kaiser prediction, $P(k,\mu) = (b_L+f\mu^2)^2P_L$, using the linear bias 
$b_L = \sqrt{P_H/P_{DM}}|_{k<0.05}$.
Panels correspond to the LRG HOD model (top left), the 
 \vmpeak sample (top right), the \mpeak sample (bottom left), and the \vmax sample (bottom right).
}
\label{fig:kaiser}
\end{center}
\end{figure*}

Although RSD effects on small scales are  the result of nonlinear growth, the distinction we make in this paper is between nonlinear growth in real space and the distortions along the line of sight due to peculiar velocities. In Fig.  \ref{fig:phod}, we removed the contribution of nonlinear growth in real space to the clustering signal in order to analyse all RSD effects as a function of scale.
 We can further isolate the effect of both nonlinear growth and nonlinear RSD effects by showing the ratios of 
 $P(k,\mu)$ to the linear theory Kaiser prediction, $P(k,\mu) = (b_L+f\mu^2)^2P_L$, using the linear bias 
$b_L = \sqrt{P_H/P_{DM}}|_{k<0.05}$ measured on large scales  $k<0.05 h$Mpc$^{-1}$. This is plotted in the upper left (right) panel in 
Fig. \ref{fig:kaiser}
for the HOD (\vmpeak) model.

In Fig. \ref{fig:kaiser} a ratio of unity indicates that the Kaiser linear theory prediction is accurate on very large scales, in agreement with previous work
\citep{Scoccimarro:2004tg, 2010PhRvD..82f3522T,2011MNRAS.410.2081J, 2012MNRAS.427L..25J, 2012ApJ...748...78K, 2015MNRAS.446...75B}.
This figure also shows the impact of nonlinear growth, increasing the ratio above unity for some $\mu$ bins, followed by a turn around/damping due to virial velocities. In all the panels of  Fig. \ref{fig:kaiser} the $\mu=0.1$ bin  (orange) is dominated by nonlinear growth and the clustering signal is not appreciably affected by RSD effects. The $\mu=0.5$ and $\mu=0.3$ data for the \vmpeak model shows greater nonlinear growth  (factor of 2 and 1.4 increase in clustering signal above the Kaiser prediction) compared to the HOD model which shows a factor of 1.7 and 1.2 for the same bins.

The lower left (right) panel of Fig. \ref{fig:kaiser} shows similar ratios for the \mpeak\ (\vmax) models.
These figures reveal an interesting feature in the $\mu=0.9$ bin, as the ratio of the anisotropic power spectrum $P(k,\mu)$  to the linear theory Kaiser prediction is close to unity on scales $k<0.4 h$Mpc$^{-1}$ for the HOD, \vmpeak\ and \mpeak\ models. This implies that on these scales the $\mu=0.9$ data is dominated by linear RSD effects which we can correctly model once we know the linear bias. 
When comparing the different models in Fig. \ref{fig:kaiser} it  can be seen that the turnaround scale, at which RSD damping effects are greater than the nonlinear growth, is different not only for each  $\mu$ bin but also for each model. These features may provide unique signatures in real galaxy data which could be used to distinguish between different models for populating halos with galaxies. We shall explore this in more detail in future work.

\subsection{Extracting the nonlinear bias from $P(k,\mu<0.2)$  \label{sec:extract_b}}

\begin{figure*}
\begin{center}
\includegraphics[height=3.in,width=3.5in]{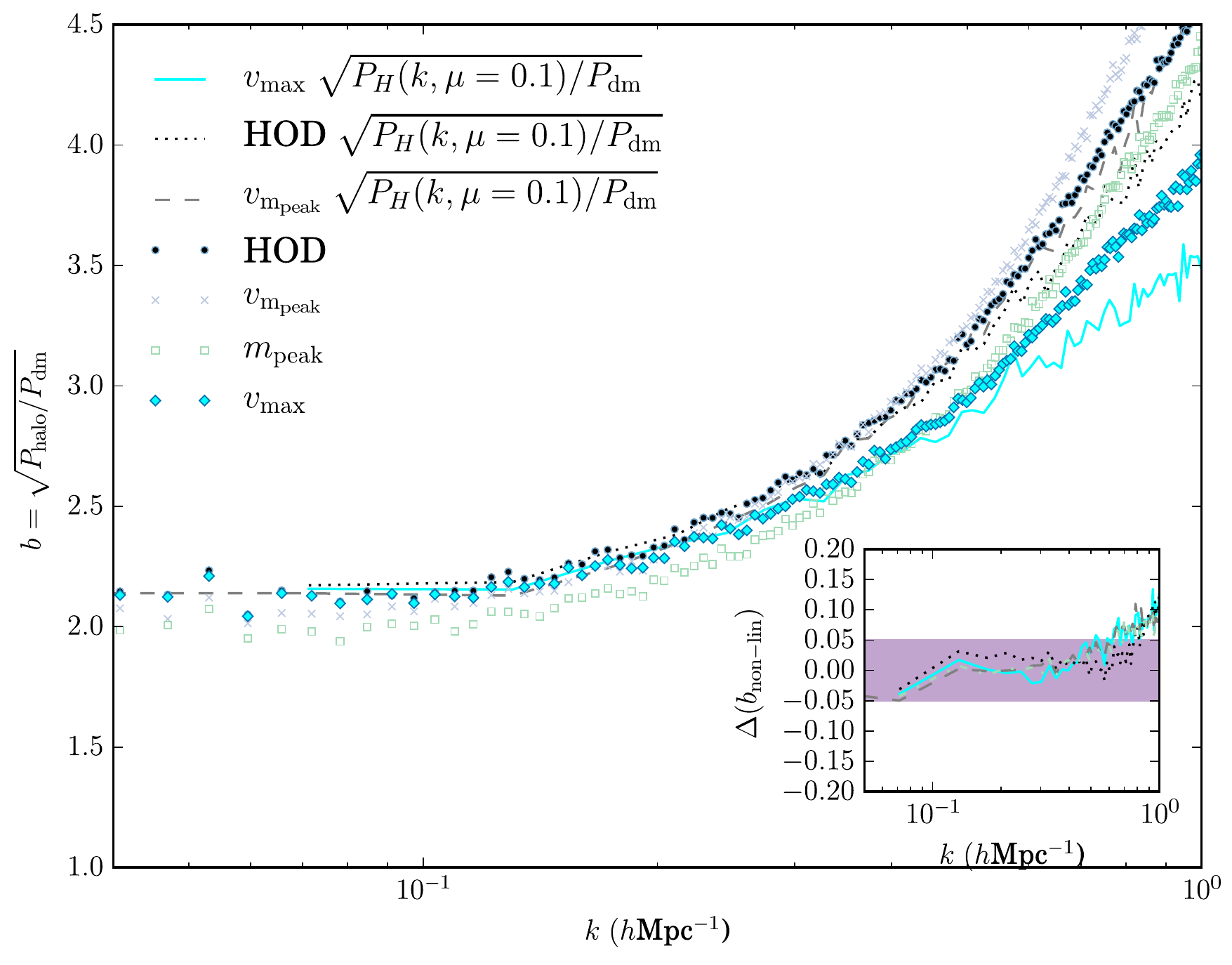}
\caption{The nonlinear bias, $b_{\rm NL} = \sqrt{P_H/P_{\rm DM, \, lin}}$, from the \mpeak (squares), \vmpeak (crosses), \vmax (diamonds) and the HOD (circles) models. The nonlinear bias measured from the 
$\mu = 0.1$ bin $\sqrt{P(k,\mu=0.1)/P_{\rm DM, \, lin}}$ is shown as a grey dashed, cyan solid and black dot dashed line for the \vmpeak, \vmax, and HOD samples respectively. We do not plot $\sqrt{P(k,\mu=0.1)/P_{\rm DM, \, lin}}$ for \mpeak\ for clarity. The inset panel shows the ratio of the nonlinear bias $b_{\rm NL}$ measured from the auto power spectra in real space to the one measured using the $\mu=0.1$ power spectrum bin. The ratio for the \mpeak model is shown as a dot-dashed line. The shaded band shows $\pm 5$\% difference in the ratio. }
\label{fig:bias}
\end{center}
\end{figure*}

Motivated by Fig. \ref{fig:phod}, which shows that the lowest $\mu=0.1$ bin is not significantly impacted by RSD effects on  large scales, in this section we test the accuracy of extracting the nonlinear bias from this $\mu$ bin and consider how this measurement might improve the RSD models for other $\mu>0.2$ bins. In Fig. \ref{fig:bias} we show the measured nonlinear bias $b_{\rm NL} = \sqrt{P_H/P_{\rm DM, \, lin}}$ for the HOD (circles), \vmax (diamonds), \mpeak (squares) and \vmpeak (crosses) models and compare this in each case with the nonlinear bias extracted from the $\mu = 0.1$ bin $\sqrt{P(k,\mu=0.1)/P_{\rm DM, \, lin}}$. Note $P_H$ is the mock galaxy power spectrum in real space for each model. The extracted nonlinear bias is shown as  a dotted black, cyan solid, and grey dashed line for the HOD, \vmax, and \vmpeak models. We do not plot $\sqrt{P(k,\mu=0.1)/P_{\rm DM, \, lin}}$ for \mpeak for clarity. The inset panel shows the ratio of the nonlinear bias $b_{\rm NL}$ measured from the auto power spectra in real space to the one measured using the $\mu=0.1$ power spectrum bin. The ratio for the \mpeak\ model is shown as a dot-dashed line. The shaded band shows $\pm 5$\% difference in the ratio. 

 From Fig. \ref{fig:bias} it is clear that the nonlinear bias inferred using the $\mu = 0.1$ power spectrum bin in redshift space is accurate to better than 5\%  over the range $k<0.6 h$Mpc$^{-1}$. Importantly, this is less than the difference in the predicted nonlinear bias between the models ( c.f. Fig. \ref{fig:pk_k_mu}), which is about 10\%. If we have an accurate forecast for the nonlinear bias in different galaxy population models and we can subsequently extract the nonlinear bias from the $\mu=0.1$ galaxy clustering data, this could potentially be used to distinguish between different methods for populating halos.

There is another advantage to using this $\mu=0.1$ data in order to measure the nonlinear bias. In the standard approach to measuring the linear growth rate from redshift-space clustering statistics, there is a degeneracy between the growth rate, $f$, and the bias. If we are able to constrain the linear bias accurately from low $\mu <0.2$ bins and use this to jointly constrain $f$ from the $\mu>0.2$ data, this might give rise to improved constraints on $f$. The results from this fitting procedure  are shown in the next section.

\begin{figure*}
\begin{center}
\includegraphics[height=3.in,width=3.2in]{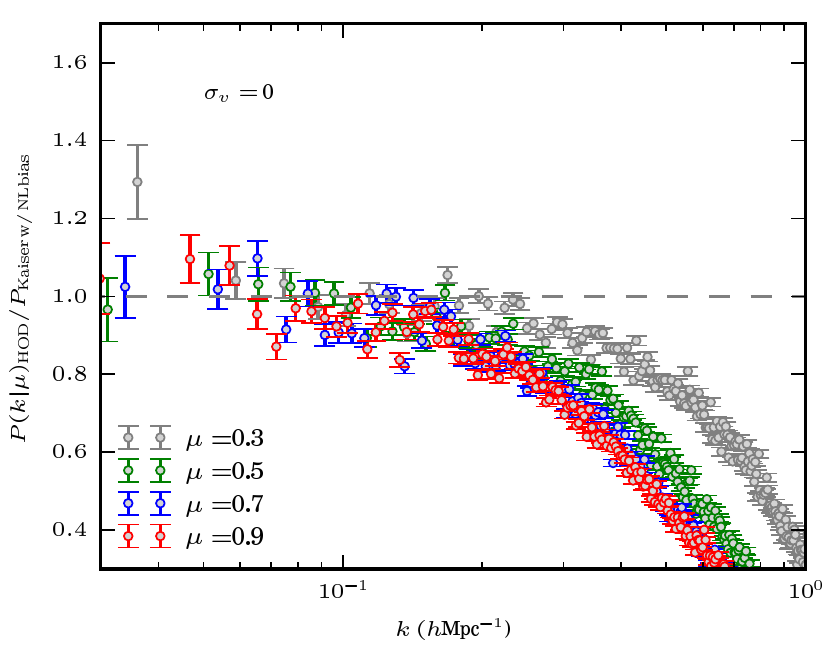}
\includegraphics[height=3.in,width=3.2in]{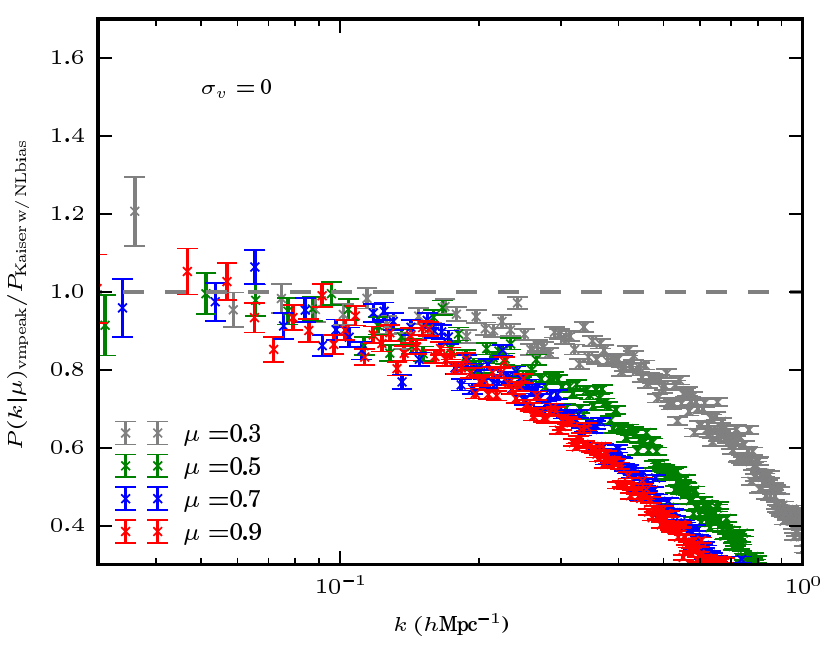}
\caption{{\em Left:} The ratio of the anisotropic redshift space power spectrum for the HOD model to the Kaiser prediction, $P(k,\mu) = (b_{\rm NL}+f\mu^2)^2P_L$ using the nonlinear bias measured from the $\mu=0.1$ simulation data.  
{\em Right:} Same ratio as in the left panel for the \vmpeak model.
}
\label{fig:kaisernlb}
\end{center}
\end{figure*}

The left (right) panel in Fig. \ref{fig:kaisernlb} shows the ratio of $P(k,\mu)$ for the HOD (\vmpeak) model
 to the Kaiser prediction,$P(k,\mu) = (b_{\rm NL}+f\mu^2)^2P_L$, where the nonlinear bias has been extracted from the $\mu = 0.1$ data.
Including the nonlinear bias in this simple way removes the nonlinear enhancements seen in Fig.  \ref{fig:kaiser}  and reveals a 
damped signal which is different for each $\mu$ bin. 
It is clear from this figure that now only nonlinear RSD damping effects are present and  modeling these with an exponential damping term seems reasonable.

\begin{figure*}
\begin{center}
\includegraphics[height=3.in,width=3.2in]{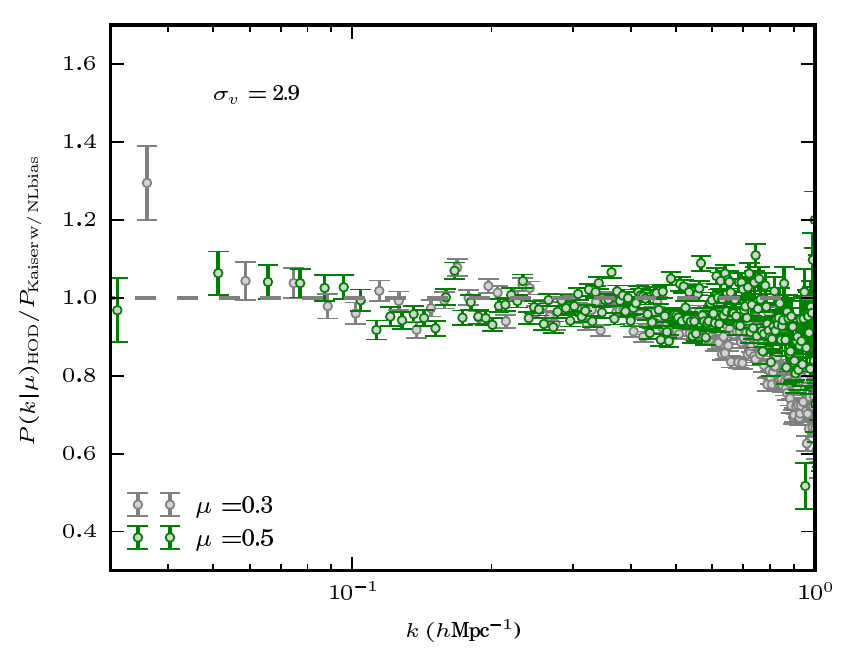}
\includegraphics[height=3.in,width=3.2in]{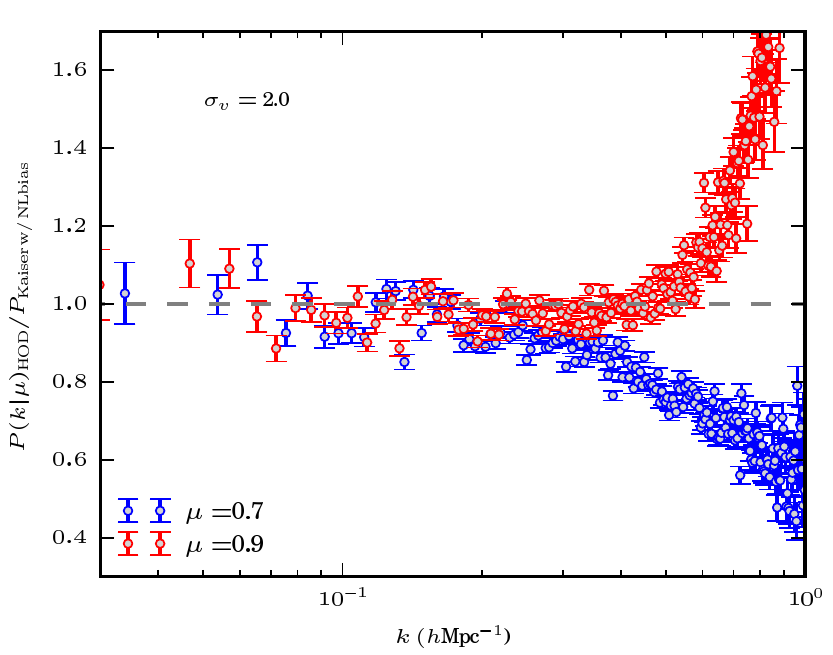}
\caption{{\em Left:} The ratio of the anisotropic redshift space power spectrum at $\mu=0.3$ and $\mu=0.5$ for the LRG HOD sample to the dispersion model prediction using the nonlinear bias extracted from the $\mu = 0.1$ data. A fixed damping factor of $\sigma_v=2.9$ has been used for each.
{\em Right:} The same ratio as in the left panel for $\mu=0.7$ and $\mu=0.9$ with a fixed damping factor $\sigma_v=2$ for each.
}
\label{fig:kaisernlbsigma}
\end{center}
\end{figure*}

As outlined in Section \ref{sec:rsd_nonlinear}, the dispersion model is an extension of Kaiser linear theory where the damping effects on small scales are taken into account,
$P(k,\mu) = (b_{\rm NL}+f\mu^2)^2P_L e^{-(f\mu\sigma_v)^2}$. Here $\sigma_v$ is the pairwise velocity dispersion which is assumed to be scale independent; it is common to treat this as a constant free parameter in the fit for $f$. 
This assumption is not correct in general, as the power spectrum in redshift space contains an exponential term which is scale dependent through the pairwise line of sight velocity difference $v_z({\bf x})-v({\bf x'})$ for galaxies at positions ${\bf x}$ and ${\bf x'}$ in real space \citep[see e.g ][]{Scoccimarro:2004tg}. 

Note  in Fig. \ref{fig:kaisernlb} $\sigma_v = 0$.

In the left  (right) panel of Fig. \ref{fig:kaisernlbsigma}, we show the ratio of the $\mu=0.3, 0.5$ ($\mu=0.7, 0.9$) 
data for the HOD model to the dispersion model
 where the nonlinear bias has been extracted from the $\mu = 0.1$ data and we have fixed $\sigma_v = 2.9$ ($\sigma_v = 2.0$) for demonstration purposes.
Previous work in fitting to either the quadrupole to monopole ratio or to the full $P(k,\mu)$ assumes a fixed $\sigma_v$ for all bins although it is clear that allowing a different free parameter for each $\mu$ bin gives a better fit to the data.
Given the complex shape and relative importance of nonlinear growth and linear and nonlinear RSD effects to
the 2D power spectrum as seen in Fig. \ref{fig:kaiser}, adding a damping term with a $\mu$-dependent 
free parameter extends the number of useful modes with which we can reliably constrain cosmology.
We present results of this fitting procedure in the next section.

\begin{figure*}
\begin{center}
\includegraphics[height=3.in,width=3.2in]{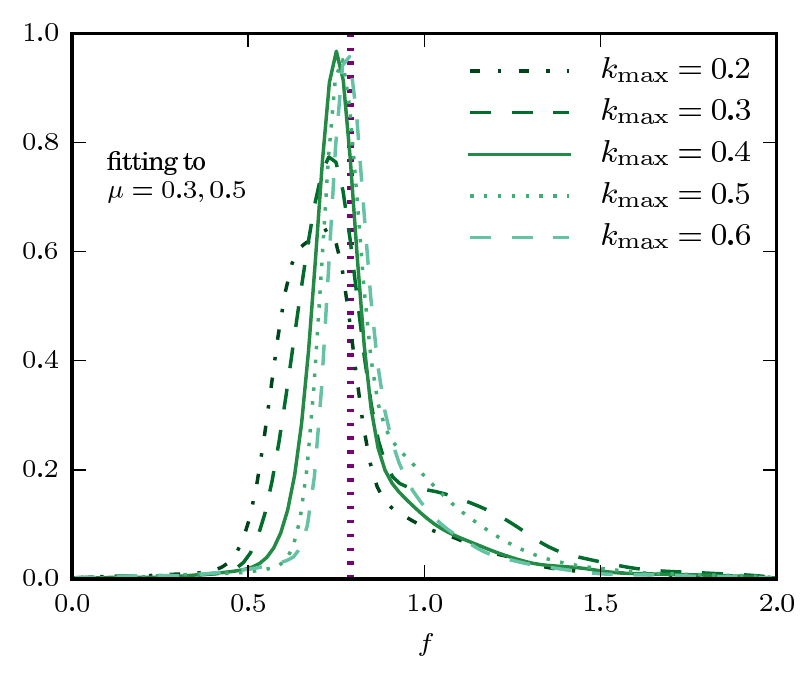}
\includegraphics[height=3.in,width=3.2in]{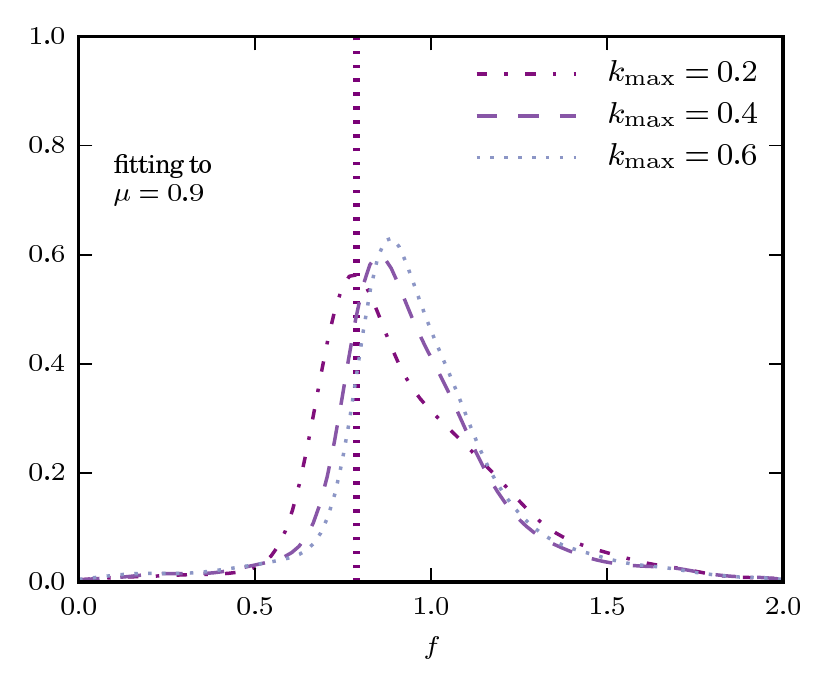}
\caption{{\em Left:}The 1D marginalized likelihood for the growth rate $f(z=0.67)$ as a function of the maximum wavenumber used in the fit, $k_{\rm max}$, fitting to $P(k,\mu)$ for the HOD sample using the $\mu=0.3$ and $\mu=0.5$ simulation data. 
{\em Right: } The 1D marginalized likelihood for $f$ as a function of $k_{\rm max}$ using the $\mu=0.9$ bins.
The $\mu=0.1$ data was used to simultaneously fit for the nonlinear bias in both panels. The actual value for the growth rate is shown as a vertical dashed line in each panel.
}
\label{fig:growthrate}
\end{center}
\end{figure*}

\begin{figure*}
\begin{center}
\includegraphics[height=4.in,width=4.5in]{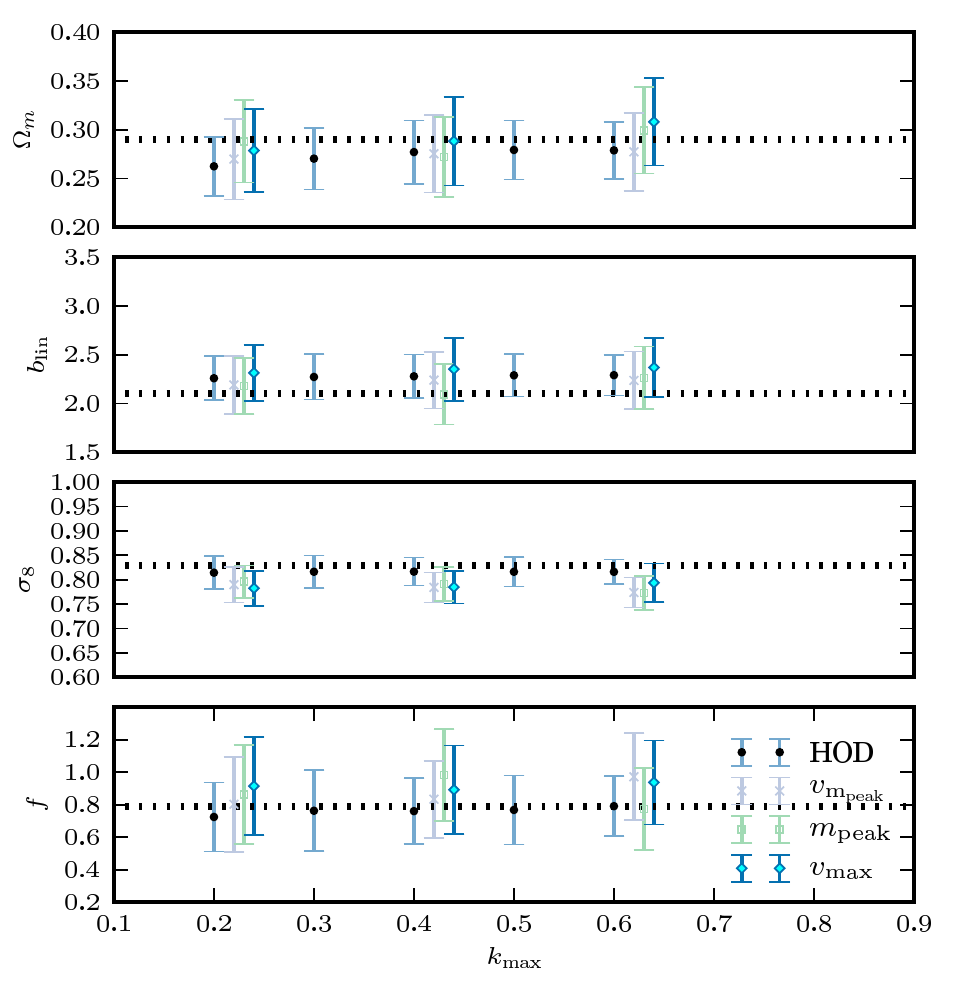}
\caption{The median and $1-\sigma$ errors for the growth rate $f(z=0.67)$, $\sigma_8$, $b_{\rm lin}$ and $\Omega_m$ as a function of the maximum wavenumber used in the fit, $k_{\rm max}$ for the HOD model (black circles). Similar results are plotted for the \vmpeak (crosses), \vmax (diamonds) and \mpeak (squares) models only at $k_{\rm max}$ = 0.2, 0.4  and 0.6 $h$Mpc$^{-1}$ for clarity. }
\label{fig:parameters}
\end{center}
\end{figure*}

\subsection{Parameter estimation \label{sec:params}}

In this section, we present simple examples of parameter fitting  to different bins in $\mu$ from the 2D redshift space $P(k,\mu)$ taking into account the effect of nonlinear growth and RSD on each. 
We present the results from a joint analysis of the $\mu=0.1$ simulation data fitting only for the nonlinear bias as given in Eq. \ref{eq:nlbias}; and fitting the $\mu=0.3$ and $\mu=0.5$ data for the linear growth rate and $\sigma_8$ using this nonlinear bias in the Dispersion model.
We shall also show results of fitting to the $\mu=0.9$ bin using a  joint fit to $\mu=0.1$ to extract the nonlinear bias. Note we fit to $\mu=0.3$ and $\mu=0.5$ using a single free parameter $\sigma_v$ and fit a separate free parameter to the $\mu=0.9$ simulation data. This approach is motivated by Fig. \ref{fig:kaisernlbsigma} where it appears that different damping factors are required for each.

In order to constrain the linear growth rate from simulated clustering measurements in redshift space we 
 use the {\sc EMCEE} ensemble sampler \citep{2013PASP..125..306F} as part of the publicly available parameter estimation code {\sc CosmoSIS} \citep{2015A&C....12...45Z}. We use 400 walkers for 400 steps which, after thinning by a factor of 2 and discarding burn-in, yields 59,000 independent samples from the posterior. We jointly fit the $\mu=0.1$ simulation data for the nonlinear bias and the $\mu=0.3$ and 0.5 data allowing the following set of parameters to vary:
 \{ $\Omega_m$, $\sigma_8$, $f$, $\sigma_v$, $b_{\rm lin}$, $Q$\}, where $b_{\rm lin}$ and $Q$ are parameters in the nonlinear bias model in Eq. \ref{eq:nlbias}.
 Our pipeline consists of first running {\rm CAMB} \citep{Lewis:2002ah} to output the linear matter power spectrum at $z=0.67$, then evaluating the log likelihood for the nonlinear bias using the $\mu=0.1$ simulation data, and finally evaluating the log likelihood for the $\mu=0.3$ and 0.5 power spectra in redshift space using the model in Eq. \ref{eq:exp} at each point in parameter space. This pipeline is repeated using the $\mu=0.9$ simulation data.

In the left (right) panel of Fig. \ref{fig:growthrate} we show the results of fitting simultaneously to the $\mu=0.1$ bin for the nonlinear bias and the $\mu=0.3, 0.5$ (0.9) data for the linear growth rate $f$ at redshift $z=0.67$ using the HOD sample and varying the maximum wavenumber used in the fit, $k_{\rm max}$. Note the same fixed 
 $\sigma_v$ parameter was assumed for both $\mu=0.3 $ and $0.5$. We have also assumed that the error on different $k-$bins are uncorrelated over the scales used in the fit. This plot shows that up to a $k_{\rm max} = 0.6$ we are able to recover a correct estimate of the linear growth rate. As the maximum wavenumber used in the fit is increased, it is clear that the overall errors on $f$ decrease and still recover the correct estimate of the growth rate without any bias. We do not extend the fit beyond $k_{\rm max} = 0.6$, as from Fig. \ref{fig:bias} this is the scale at which we expect the estimate of the nonlinear bias from the $\mu=0.1$ bin to no longer be accurate. We find that the constraints on $f$ using 
 $\mu=0.9$ are not as accurate down to small scales (right panel of Fig. \ref{fig:growthrate}) as we recover a mean growth rate of $f=0.94 \pm 0.25$ which is slightly high. However, this is still consistent with the actual value of $f(z=0.67) = 0.79$.

The median values with $1\sigma$ errors for $\Omega_m$, $b_{\rm lin}$, $\sigma_8$ and $f$ are shown in Fig. \ref{fig:parameters} for the HOD (circles) model at $k_{\rm max} = 0.2, 0.3, 0.4,0.5 $ and $0.6$.
Similar results for the \vmpeak (crosses), \mpeak (squares) and \vmax (diamonds) models are shown at $k_{\rm max} = 0.2, 0.4$, and $0.6$. These results give an approximate error on $f$ of 26 (22)\% to $k_{\rm max}=0.4 (0.6)$ for the HOD model from clustering data only. Reassuringly we find that the constraints on these cosmological parameters are not very dependent on our choice of galaxy model; we are able to recover unbiased cosmological parameters for each of the models despite the fact that these models have different redshift-space clustering signals. 

This result greatly outperforms the application of the dispersion model to the multipole moments in previous studies both in the range of scales used and in the accuracy with which the growth rate is recovered.  \citet{2015MNRAS.447..234W} recently showed that several so called ``streaming" models of redshift-space distortions fail to recover a correct value for the growth rate and are significantly biased on scales $r<25 h^{-1}$Mpc ($k_{\rm max}>0.25 h$Mpc$^{-1}$) when fitting to the monopole-to-quadrupole ratio. Previous galaxy surveys have also been limited to very large scales $k_{\rm max}<0.2h$Mpc$^{-1}$, due to a breakdown of our theoretical models for bias and redshift-space distortions\citep{2008Natur.451..541G,2011MNRAS.415.2876B,2012MNRAS.423.3430B}. In addition, forecasts for future surveys generally  include CMB (Planck \citep{2014A&A...571A...1P}) and weak gravitational lensing (DES, \citealt{2015arXiv150705598B} and LSST, \citealt{2009arXiv0912.0201L}) constraints in order to reduce the uncertainty on the linear growth rate
\citep[e.g.][]{2014JCAP05023F}, whereas in this work we present constraints using clustering data only.
In this paper we have considered a very simple RSD model to illustrate the concept of disentangling redshift space effects from nonlinear bias using the $\mu$ dependence in the 2D power spectrum. We shall investigate how these parameter constraints might improve using the same method but with more robust, physically motivated models for both the bias and redshift-space distortions in a future study.

Note our method in this section is different in several ways to the commonly used approach of analyzing the projected two point correlation, $w_p(r_p)$, \citep{1983ApJ...267..465D} where the RSD effects are minimized by integrating the redshift space correlation function, $\xi(r_p,\pi)$ along the line of sight, $\pi$ \citep[see e.g.][]{2015MNRAS.446..578G}.
First, using the projected correlation function does reduce the dependence RSD but does not remove their effect e.g. \citet{2010MNRAS.407..520N} showed that RSD still impact the clustering signal by the apparent movement of galaxies into or out of the sample.
Secondly, $w_p(r_p)$ does not directly allow us to measure the real space correlation function of the galaxies and so we cannot simply determine the bias.
It is possible to estimate $\xi(r)$ in real space by inverting the equation for $w_p(r_p)$ \citep{1992MNRAS.258..134S} but as shown in Fig 1. \citet{2009MNRAS.393.1183C} this is not precise and the results depend heavily on the upper limit of the integral for $w_p(r_p)$. 
Our approach is also different to the clustering wedges method of
\citet{2013MNRAS.433.1202S} which splits the redshift space correlation function into two bins but does not attempt to separate nonlinear bias and RSD effects.
In this paper we show that the $\mu<0.2$ modes are independent of RSD effects at $k<0.6h^{-1}$Mpc 
and that the nonlinear bias can be reconstructed to an accuracy of $<5$\% by fitting to this bin.

\section{Summary \& Conclusions 
{\label{sec:conc}}}

In this paper we analyze the 2D power spectrum in redshift space using a 1 \hGpc cubed volume simulated with 1 trillion particles, from the {\em Dark Sky} simulation series. Using several different models for populating halos with galaxies we generate mock LRG samples at a redshift of $z=0.67$, with number densities relevant for future spectroscopic galaxy surveys. We consider a halo occupation distribution as well as subhalo abundance matching models, which assign galaxies to halos based on halo/subhalo properties of \vmax (the maximum circular velocity at the present time), \mpeak (the maximum mass the halo or subhalo has ever had throughout its merger history) and \vmpeak (the maximum circular velocity when the halo or subhalo has achieved its \mpeak).  In this work we present the redshift-space clustering signal of these models for the first time. The high mass and force resolution of the Dark Sky Gpc simulations is essential for resolving substructure within virialized haloes needed for subhalo abundance matching, as well as accurately measuring the nonlinear growth in both the velocity and density fields.

All of the galaxy models predict different satellite fractions and clustering amplitudes in real space. As a result we find differences in the linear bias on large scales ($\sim 10$\%) which increase on small scales due to different satellite fractions (we also consider models with different levels of assembly bias). We find a clear trend of increased velocity dispersion and damping of the redshift-space clustering signal on quasi-linear to nonlinear scales which follows from the satellite fraction in each model. We find the model with the largest satellite fraction, \vmpeak, has larger RSD effects then the HOD, \mpeak and \vmax models when we examine both the quadrupole to monopole ratio and power spectra along certain angles with respect to the line of sight  $\mu = \cos(k_z/|k|) >0.2$.

One of the key findings in this paper is that different bins in $\mu$ can be used to isolate the impact of nonlinear growth and RSD effects.  One interesting outcome is that the lowest $\mu <0.2$ bins are unaffected by RSD effects and so it is possible to extract a good estimate of the non-linear bias at $k<0.6 h$Mpc$^{-1}$. Our analysis of individual $\mu $ bins  also reveals some interesting physical effects; we find a prominent turnaround scale, at which RSD damping effects are greater than the nonlinear growth, which is different not only for each $\mu$ bin but also differs for each galaxy model. These features may provide unique signatures which could be used to distinguish between the  different models for the galaxy--halo connection.
 
Using a simple dispersion model we present results from a joint analysis of the $\mu$ = 0.1 simulation data fitting for the nonlinear bias and the $\mu$ = 0.3 and $\mu$ = 0.5 data for the linear growth rate, the linear bias, and $\sigma_8$. We can recover 
 $f$ to an accuracy of $\sim 26 (22)$\% to  
$k_{\rm max}< 0.4 (0.6) h$Mpc$^{-1}$ from the HOD model.
This result greatly outperforms the application of the dispersion model in previous studies  both in the range of scales used and the accuracy with which 
the growth rate is recovered. 
To put these results in context, \citet{2015MNRAS.447..234W} recently showed that several commonly used models of redshift-space distortions fail to recover a correct value for the growth rate and are significantly biased on scales $k_{\rm max}>0.25 h$Mpc$^{-1}$. Previous galaxy surveys have also been limited to very large scales $k_{\rm max}<0.2h$Mpc$^{-1}$ due to a breakdown of our theoretical models for bias and redshift-space distortions 
\citep{2008Natur.451..541G,2011MNRAS.415.2876B,2012MNRAS.423.3430B}. In this work we present constraints using clustering data only whereas forecasts for future surveys generally  include CMB  and weak gravitational lensing constraints in order to reduce the uncertainty on cosmological parameters
\citep[e.g.][]{2014JCAP05023F}.
 In a future study we shall investigate how these parameter constraints might improve using more robust, physically motivated models for redshift-space clustering and combining data sets.

We find that the constraints on these cosmological parameters are not particularly sensitive to the galaxy formation model used. This result is reassuring if we are interested in unbiased cosmological constraints as there are many variations in the way we can model the connection between galaxies and dark matter.  The results presented here show the wealth of information that is available in the full 2-D redshift-space power spectrum. In a follow-up analysis we shall explore in detail how several features in $P(k,\mu)$, for example the transition from nonlinear growth to RSD damping, may also be used to distinguish galaxy models and shed light on the galaxy--halo connection.

\section{Acknowledgements}
This research made use of the Dark Sky Simulations, which were produced using an INCITE 2014 allocation (M. Warren et al) on the Oak Ridge Leadership Computing Facility at Oak Ridge National Laboratory.  We are grateful to Matt Turk and the rest of our Dark Sky collaborators.  We also thank Yao-Yuan Mao for running the merger trees, and to Benjamin Lehmann for assistance with galaxy catalog production.  We are grateful to the University of Chicago Research Computing Center and we thank the scientific computing team at SLAC for their support related to hosting data through the darksky server.  We thank Eduardo Rozo for a careful reading of the first draft, and  Carlton Baugh, Scott Dodelson and Andrey Kratsov for useful discussions. EJ is supported by Fermi Research Alliance, LLC under the U.S. Department of Energy under contract No. DE-AC02-07CH11359. RHW received support from the U.S. Department of Energy under contract number DE-AC02-76SF00515.

\bibliographystyle{mn2e}
\bibliography{thebibliography}

\end{document}